\documentclass[]{aastex61}
\usepackage{epstopdf}
\usepackage{tikz}

\newcommand\aastex{AAS\TeX}

\received{xx}
\revised{xx}
\accepted{xx}
\submitjournal{ApJ}

\shorttitle{\aastex\ sample article}
\shortauthors{Cai}

\begin{document}

\title{Upward Overshooting in Turbulent Compressible Convection. III. Calibrate Parameters for One-dimensional Reynolds Stress Model}

\correspondingauthor{Tao Cai}
\email{tcai@must.edu.mo, caitao7@mail.sysu.edu.cn}

\author[0000-0003-3431-8570]{Tao Cai}
\affil{State Key Laboratory of Lunar and Planetary Sciences, Macau University of Science and Technology, Macau, People's Republic of China}
\affil{School of Mathematics, Sun Yat-sen University, No. 135 Xingang Xi Road, Guangzhou, 510275, People's Republic of China}



\begin{abstract}
In this paper, we calibrate the coefficients for the one-dimensional Reynolds stress model with the data generated from the three-dimensional numerical simulations of upward overshooting in turbulent compressible convection. It has been found that the calibrated convective and isotropic coefficients are almost the same as those calibrated in the pure convection zone. However, the calibrated diffusive coefficients differ significantly from those calibrated in the pure convection zone. We suspect that the diffusive effect induced by the boundary is stronger than by the adjacent stable zone. We have checked the validity of the downgradient approximation. We find that the prediction of the downgradient approximation on the third-order moments is unsatisfactory. However, the prediction on their derivatives is much better. It explains why the performance of the Reynolds stress model is reasonable in application to the real stars. With the calibrated coefficients, we have solved the full set of nonlocal turbulent equations on Reynolds stress model. We find that the Reynolds stress model has successfully produced the thermal adjustment layer and turbulent dissipation layer, which were identified in the three-dimensional numerical simulations. We suggest to use the inflection point of the auto-correlation of temperature perturbation and the P\'eclet number as the indicators on measuring the extents of the thermal adjustment layer and turbulent dissipation layer, respectively. This result may offer a practical guidance on the application of the Reynolds stress model in 1D stellar structure and evolution models.
\end{abstract}

\keywords{convection --- methods: numerical --- hydrodynamics --- stars: interiors}



\section{Introduction} \label{sec:intro}
Core overshooting is an important physical process for the evolutions of the intermediate and massive stars. The duration of the hydrogen and helium burning stage can be considerably affected, since more fuels are supplied through the extra mixing induced by the overshooting process. Many studies have confirmed that core overshooting is needed to explain the observational results. For example, the study of double-lined eclipsing binaries indicates that the overshooting parameter is non-zero for stars more massive than $1.2M_{\odot}$ \citep{claret2017dependence,claret2018dependence,claret2019dependence}. Similarly, it has been found that including the core overshooting can significantly decrease the discrepancies between the theoretical and observed apsidal motion rates in double-lined eclipsing binaries \citep{claret2010apsidal,claret2019updating}. Recently, evidence from asteroseismology also reveals that core overshooting is required to reproduce the seismic observations of the Kepler data \citep{deheuvels2016measuring}. While the mixing length theory \citep{bohm1958wasserstoffkonvektionszone} has been widely used in the treatment of stellar convection, uncertainty arises in its application of dealing with the energy transportation and material mixing in the overshooting zone \citep{renzini1987some}. To overcome this problem, advanced nonlocal Reynolds stress models (RSMs) have been developed \citep{xiong1981statistical,xiong1997nonlocal,canuto1992turbulent,canuto1998stellar,li2012k,li2017applications}. Despite the complexity, applications of RSMs to stellar convection and overshooting have shown promising results \citep{xiong2001structure,xiong2010non,kupka2002star,montgomery2004white,li2018convective,guo2019convective}. The equations of RSMs are incomplete unless closure relations are assumed for the higher order moments. These approximated closure relations induce truncation errors. The validity of these approximations needs to be examined. In addition, the RSMs usually contains undetermined coefficients. The values of these coefficients need to be calibrated.

Numerical simulations provide useful insights into the turbulent convection and overshooting in stars. Over the past years, much efforts \citep{chan1989turbulent,grossman1996theory,kupka2006effects,kupka2007probinga,kupka2007probingb,kupka2007probingc,garaud2010model,arnett2015beyond,cai2018numerical} were devoted to bridge the gap between the numerical simulations and the theoretical models. In these works, different closure models were tested by the three-dimensional numerical simulation data. Although some of the testing results have promising implications, their applications to stellar convection still remain challenging. In the calculation of stellar models, the equations of RSMs have to be solved together with the thermal structure equations. Previous attempts have shown that numerical instability occurs in most cases when the closure models of higher order moments are involved \citep{grossman1996theory,kupka2007probingb,cai2018numerical}. However, the calculation of RSMs derived from the downgradient approximations (DGAs), seems to be numerically stable \citep{cai2018numerical}. In our previous calculation \citep{cai2018numerical}, we have calibrated the coefficients of Xiong's 1D RSM by the three-dimensional (3D) simulation data of efficient turbulent convection in a pure convection zone. The convective and isotropic coefficients are calibrated by a local steady approximation in the convection zone. The diffusive coefficients are calibrated by a power law approximation derived from the boundary effect. In the stellar interiors, convectively stable zones are usually attached to the convection zone. The diffusive effect induced by boundary effect might be different from that induced by the attached stable zone. In addition, the important overshooting process needs to be investigated.

In recent years, much attention has been paid to the numerical simulations of the turbulent convection and overshooting \citep{brummell2002penetration,hotta2017solar,brun2017differential,kapyla2017extended,kapyla2019overshooting,korre2019convective}. Most of them studied the downward overshooting below a convection zone. \citet{browning2004simulations} studied the upward overshooting in massive stars, and they mainly focused on the effects of rotation on upward overshooting. In a previous paper \citep{cai2020upward}, we studied the upward overshooting above a convection zone by numerical simulations. We have found that the upper convectively stable zone can be separated into three layers: the thermal adjustment layer, the turbulent dissipation layer, and the thermal dissipation layer. The theoretical work of \citet{zahn1991convective} explained the difference between penetration (nearly adiabatic) and overshooting (non-adiabatic) from a physical point of view. In his work, the penetrative and overshooting zones were also called nearly adiabatic layer and thermal adjustment layer, respectively. Based on a RSM, \citet{zhang2012turbulent} predicted that an additional turbulent mixing layer exists. Our numerical result \citep{cai2020upward} has shown a remarkable qualitatively agreement with the theoretical prediction of the 1D RSM in \citet{zhang2012turbulent}.  In this paper, we take a further step to compare the 3D simulations with 1D RSM quantitatively. Specifically, we mainly consider the following questions. First, we calibrate the coefficients for the 1D RSM. Second, we test the validity of the downgradient approximation. Third, we link the extent of overshooting distance with physical indicators.

\section{The Model} \label{sec:model}
\subsection{The 1D nonlocal RSM}
In a previous paper, we have compared Xiong's 1D nonlocal RSM with the 3D simulations of efficient turbulent convection in the Cartesian geometry \citep{cai2018numerical}. Here, we extend the previous research to compare the 1D nonlocal RSM with the 3D simulations on the upward overshooting of turbulent convection \citep{cai2020upward}. For the convenience of illustration, we list the downgradient form of Xiong's RSM here again. In the Cartesian coordinates, the dowgradient form of Xiong's RSM model is \citep{cai2018numerical}:
\begin{eqnarray}
\frac{\partial \overline{u^2}}{\partial t} -\frac{1}{\rho}\frac{\partial}{\partial z}[\frac{\sqrt{3}}{4}c_{2,w^2}\frac{P}{g} (\overline{w^2})^{1/2} \frac{\partial}{\partial z}\overline{u^2}]-\frac{2}{3}{\beta}g (\overline{w\frac{\theta}{T}}) +\frac{4\eta_{e}}{\sqrt{3}}\frac{\rho g}{c_{1,w^2}P}(\overline{u^{2}})^{3/2}&=&0~,\\
\frac{\partial (\overline{\frac{\theta^2}{T^2}})}{\partial t}-\frac{1}{\rho}\frac{\partial}{\partial z}[\frac{\sqrt{3}}{4}c_{2,\theta^2}\frac{P}{g} (\overline{w^2})^{1/2} \frac{\partial}{\partial z}\overline{\frac{\theta^2}{T^2}}] +2 (\frac{\partial \ln T}{\partial z}-\nabla_{ad}\frac{ \partial \ln P}{\partial z})(\overline{w\frac{\theta}{T}})&&\\\nonumber
+2\sqrt{3}\eta_{e}\frac{\rho g}{c_{1,\theta^2}P}[(\overline{u^2})^{1/2}+u_{c}](\overline{\frac{\theta^2}{T^2}})&=&0~,\\
\frac{\partial (\overline{w\frac{\theta}{T}})}{\partial t}-\frac{1}{\rho}\frac{\partial}{\partial z}[\frac{\sqrt{3}}{4}c_{2,w\theta}\frac{P}{g} (\overline{w^2})^{1/2} \frac{\partial}{\partial z}\overline{w\frac{\theta}{T}}]+(\frac{\partial \ln T}{\partial z}-\nabla_{ad}\frac{\partial \ln P}{\partial z})(\overline{w^2})-{\beta}g (\overline{\frac{\theta^2}{T^2}})&&\\\nonumber
+\sqrt{3}\eta_{e}\frac{\rho g}{c_{1,w\theta}P}[3(\overline{u^2})^{1/2}+u_{c}](\frac{\overline{u^2}}{\overline{w^2}})^{1/2}(\overline{w\frac{\theta}{T}})&=&0~,\\
\frac{\partial \overline{w^2}}{\partial t}-\frac{1}{\rho}\frac{\partial}{\partial z}[\frac{\sqrt{3}}{4}c_{2,w^2}\frac{P}{g} (\overline{w^2})^{1/2} \frac{\partial}{\partial z}\overline{w^2}]-2{\beta}g (\overline{w\frac{\theta}{T}}) +\frac{4\eta_{e}}{\sqrt{3}}\frac{\rho g}{c_{1,w^2}P}(\overline{u^2})^{1/2}[(1+c_{3})\overline{w^2}-c_{3}\overline{u^2}]&=&0~,
\end{eqnarray}
where the symbol overline represents the temporal and horizontal average of the quantity; $\overline{u^2}$, $\overline{w^2}$, $\overline{w\frac{\theta}{T}}$,and $\overline{\frac{\theta^2}{T^2}}$ are the auto- and cross-correlations of velocity and temperature perturbations; $u$ is the isotropic part of turbulent velocity; $w$ is the vertical velocity; $\theta$ is the temperature perturbation; $P$ is the pressure; $T$ is the temperature; $\rho$ is the density; $\beta$ is the expansion coefficient of gas; $\eta_{e}=0.45$ is the Heisenberg eddy coupling constant; $\nabla_{ad}$ is the adiabatic temperature gradient; $u_c=\frac{9}{4}\frac{\kappa}{c_{1,w\theta}\rho c_{p}H_{p}}$; $\kappa$ is the conductivity; $c_{1,w^2},c_{1,\theta^2},c_{1,w\theta}$ are the convective coefficients; $c_{2,w^2},c_{2,\theta^2},c_{2,w\theta}$ are the diffusive coefficients; and $c_{3}$ is the isotropic coefficient.

Removing the time derivative terms and diffusive terms, we obtain the local steady solution of the above equations \citet{cai2018numerical}:
\begin{eqnarray}
\overline{u^{2}}&=&\frac{1}{6\eta_{e}^{2}}(\frac{3+c_{3}}{1+c_{3}})^{1/2} [c_{1,w\theta}\frac{(\frac{3+c_{3}}{1+c_{3}})(1+\frac{u_{c}}{(\overline{u^2})^{1/2}})c_{1,w^2}+2c_{1,\theta^2}}{(1+\frac{u_{c}}{(\overline{u^2})^{1/2}})+2}] (\frac{{\beta}{P}}{{\rho}})(1+\frac{u_{c}}{(\overline{u^2})^{1/2}})^{-1}(\nabla-\nabla_{ad})~,\label{eq:u2}\\
\overline{\frac{\theta^2}{T^2}}&=&\frac{1}{3\eta_{e}^2}(\frac{3+c_{3}}{1+c_{3}})^{1/2}\frac{c_{1,\theta^2}}{c_{1,w^2}}[c_{1,w\theta}\frac{(\frac{3+c_{3}}{1+c_{3}})(1+\frac{u_{c}}{(\overline{u^2})^{1/2}})c_{1,w^2}+2c_{1,\theta^2}}{(1+\frac{u_{c}}{(\overline{u^2})^{1/2}})+2}](1+\frac{u_{c}}{(\overline{u^2})^{1/2}})^{-2}(\nabla-\nabla_{ad})^{2}~,\\
\overline{w\frac{\theta}{T}}&=&\frac{\sqrt{2}}{6\eta_{e}^{2}}(\frac{3+c_{3}}{1+c_{3}})^{3/4}\frac{1}{c_{1,w^2}}[c_{1,w\theta}\frac{(\frac{3+c_{3}}{1+c_{3}})(1+\frac{u_{c}}{(\overline{u^2})^{1/2}})c_{1,w^2}+2c_{1,\theta^2}}{(1+\frac{u_{c}}{(\overline{u^2})^{1/2}})+2}]^{3/2}(\frac{{\beta}{P}}{{\rho}})^{1/2}(1+\frac{u_{c}}{(\overline{u^2})^{1/2}})^{-3/2}(\nabla-\nabla_{ad})^{3/2}~,\\
\overline{w^2}&=&\frac{3+c_{3}}{1+c_{3}}\overline{u^2}~,\label{eq:w2}
\end{eqnarray}
in the convectively unstable zone ($\nabla-\nabla_{ad}>0$), where $\nabla=\partial \ln T/\ln P$ is the temperature gradient. With this local steady solution, we can calibrate the convective coefficients $c_{1,.}$ and isotropic coefficient $c_{3}$ \citep{cai2018numerical}. In \citet{cai2018numerical}, the diffusive coefficients $c_{2,.}$ were calibrated by assuming an asymptotic power law solution near the top boundary. For the turbulent convection with an upward convectively stable zone, these diffusive coefficients can be calibrated through a more direct method. In the downgradient approximation (DGA), the third-order moments (TOMs) are assumed to be correlated with the second-order moments (SOMs):
\begin{eqnarray}
\overline{wu^2}&\approx& -\frac{\sqrt{3}}{4}c_{2,w^2}\frac{P}{\rho g} (\overline{w^2})^{1/2} \frac{\partial}{\partial z}\overline{u^2}~,\label{eq:wu2}\\
\overline{w\frac{{\theta^2}}{T^2}}&\approx& -\frac{\sqrt{3}}{4}c_{2,\theta^2}\frac{P}{\rho g} (\overline{w^2})^{1/2} \frac{\partial}{\partial z}\overline{\frac{\theta^2}{T^2}}~,\label{eq:wt2}\\
\overline{w^2\frac{{\theta}}{T}}&\approx& -\frac{\sqrt{3}}{4}c_{2,w\theta}\frac{P}{\rho g} (\overline{w^2})^{1/2} \frac{\partial}{\partial z}\overline{w\frac{\theta}{T}}~,\\
\overline{w^3}&\approx& -\frac{\sqrt{3}}{4}c_{2,w^2}\frac{P}{\rho g} (\overline{w^2})^{1/2} \frac{\partial}{\partial z}\overline{w^2}~.\label{eq:w3}
\end{eqnarray}
Given the TOMs and SOMs, the diffusive coefficients can be easily calibrated by the above relations. Previous calculation on RSM shows that the convective effect dominates the diffusive effect in the convectively unstable zone (see figs 13-14 in \citet{cai2014numerical}). The diffusive term mainly plays the role in the overshooting zone near the unstable/stable interface. Thus it is unnecessary to include the convectively unstable zone when calibrating these diffusive coefficients. In this paper, we use the data in the convectively stable zone ($\nabla-\nabla_{ad}<0$) to calibrate $c_{2,.}$.

\subsection{The 3D simulation data}
We calibrate the coefficients $c_{1,.}, c_{2,.}, c_{3}$ of the above 1D model, by using 3D data from the simulations of the upward overshooting of the turbulent convection \citep{cai2020upward}. In the 3D simulation, the initial thermal background structure is assumed in a piecewise linear polytropic state (the temperature structure is piecewise linear but the heat conductivity is piecewise constant):
\begin{eqnarray}
T/T_{*} &=& 1+\eta_{i} (1-z)~, \\
\rho/\rho_{*} &=& (T/T_{*})^{m_{i}}~, \\
p/p_{*} &=& (T/T_{*})^{m_{i}+1}~,
\end{eqnarray}
where the subscript $*$ represents the value at the interface; the subscript $i\in\{1,2\}$ is the layer index; $\eta_{i}$ is the thickness parameter; $m_{i}$ is the polytropic index; $z$ is the depth from the bottom. In our settings, we choose the adiabatic polytropic index $m_{ad}=1.5$, the polytropic index $m_{1}=1.0$ in the layer 1 ($0\leq z \leq 1$), and $m_{2}=m_{ad}+S(m_{ad}-m_{1})$ in the layer 2 ($1<z\leq 1.5$). As a result, the layer 1 is convectively unstable and the layer 2 is convectively stable. The gravitational acceleration $g=(m_{i}+1)\eta_{i}=F_{tot}(m_{i}+1)/\kappa_{i}$ is kept constant throughout the computational domain $0\leq z\leq 1.5$, where $F_{tot}$ is the total flux and $\kappa$ is the conductivity. In all the simulations cases, we set $g=4$ and $\eta_{1}=2$. Given $F_{tot}$ and $\kappa_{1}$, the parameters $m_{2}$ and $\kappa_{2}$ can be deduced from the above equation. We have run a total of 13 cases based on this initial structure. The parameters used in the simulations are listed in table \ref{tab:table1}. In this table, ${\rm Pr}(z)=c_{p}\mu/\kappa_{i}$ is the Prandtl number, ${\rm Re}(z)=\rho v'' L_{1z}/\mu$ is the Reynolds number, and ${\rm Pe}(z)={\rm Re}(z)c_{p}\mu /\kappa_{i}$ is the P\'eclet number. Here $c_{p}=2.5$ is the heat capacity at constant pressure; $\mu$ is the dynamic viscosity; $v''$ is the averaged root mean square velocity in layer 1; and $L_{1z}=1$ is the depth of layer 1.

\subsection{The measurement of overshooting distances}
\begin{figure}
\plotone{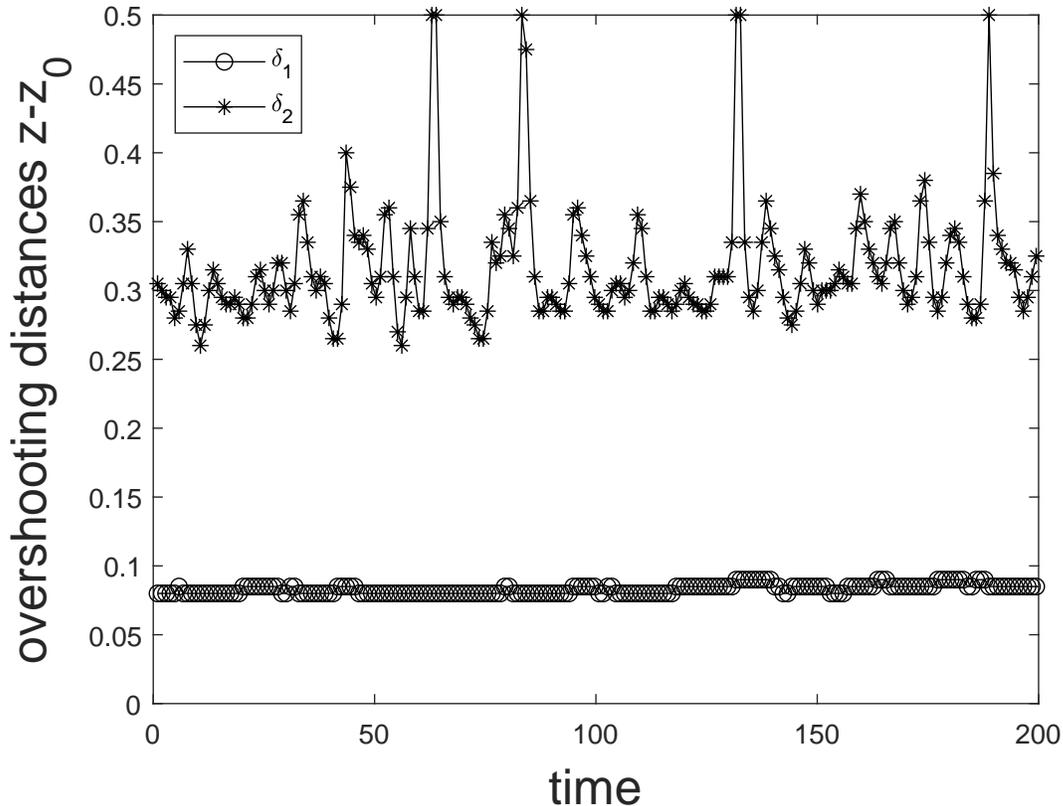}
\caption{Time variation of overshooting distances $z-z_{0}$. $z_{0}=1$ is the location of the interface between convectively unstable and stable zones. $\delta_{1}$ is depth of the thermal adjustment layer. $\delta_{2}$ is the total depth of the thermal adjustment layer and the turbulent dissipation layer. \label{fig:f1}}
\end{figure}

In \citet{chan2010overshooting}, it has been suggested to use the zeros of the vertical velocity correlation coefficients (with the vertical velocity at the convectively stable/unstable boundary) as the proxies for the measurement of overshooting distances. Following this work, we have separated the convectively stable zone of upward overshooting into three layers: the thermal adjustment layer, the turbulent dissipation layer, and the thermal dissipation layer \citep{cai2020upward}. The thermal adjustment layer and the turbulent dissipation layer are separated by the first zero point of the correlation coefficient; and the turbulent dissipation layer and the thermal dissipation layer are separated by the second zero point. The terminology `thermal adjustment layer' used here is a little bit different from what defined in \citet{zahn1991convective}. In \citet{zahn1991convective}, `thermal adjustment layer' denotes the layer where the mixing is active and the thermal structure is partially (non-adiabatically) adjusted. He set the upper boundary of this layer at the position where P\'eclet number is unity. In our work, we also found that the mixing is active and the thermal structure is partially adjusted in this layer. However, the location of the upper boundary is different. We found that the P\'eclet number at the upper boundary of this layer is significantly larger than unity. Apart from the work of \citet{zahn1991convective}, we define the upper boundary of the `turbulent dissipation layer' at the location where the P\'eclet number is unity. \citet{pratt2017extreme} have argued that averages are misleading when assessing overshooting distance. Instead, they used the statistical probability density function of extreme events to make the assessment. As mentioned in \citet{korre2019convective}, the correlation coefficients of vertical velocity tend to capture the extreme events. Thus we believe that our work shares some similarities with the work of \citet{pratt2017extreme}. Fig.~\ref{fig:f1} shows the time variation of the overshooting distance measured by the zeros of velocity correlations. It clearly shows the difference of the two distinct layers: a shallow thermal adjustment layer where convective plumes penetrate frequently; and a deeper turbulent dissipation layer where convective plumes penetrate intermittently. This result is consistent with that obtained in \citet{pratt2017extreme}.

\section{Results}
\subsection{Calibrated coefficients for 1D RSM with 3D data}
Now we calibrate the coefficients for 1D RSM. The coefficients $c_{1,.}$ and $c_{3}$ are calibrated by the eqs.(\ref{eq:u2}-\ref{eq:w2}) with the data in layer 1. This local steady solution is obtained by ignoring the diffusive terms, thus it is necessary to diminish the effect from diffusion. For this reason, the data at the top of the convection zone within $z\in(0.7,1.0)$ is excluded. Similarly, the data in the region $z\in(0,0.4)$ is excluded to avoid the undesirable boundary effect \citep{cai2018numerical}. As a result, the turbulent coefficients $c_{1,.}$ are calibrated with the data in the region $z\in (0.4,0.7)$. The isotropic coefficient $c_{3}$ is calibrated by the eq.(\ref{eq:w2}). However, an additional constraint $1\leq \overline{w^2}/\overline{u^2}\leq 3$ must be satisfied to make sure that $c_{3}$ is non-negative \citep{cai2018numerical}. It is not guaranteed that this constraint could always be satisfied in the chosen region $z\in (0.4,0.7)$. Thus the calibration of $c_{3}$ needs special treatment. In our previous work \citep{cai2018numerical}, we calibrate $c_{3}$ by using the data point at the location (within the convection zone) where $\overline{w^2}/\overline{u^2}$ achieves the maximum value. Here we follow the same strategy in this paper.

We calibrate the diffusive coefficients $c_{2,.}$ by eqs.(\ref{eq:wt2}-\ref{eq:w3}) with the 3D data. In the convectively unstable zone, the convective effect dominates the diffusive effect. As the diffusive term only plays a minor role in this region, it is unnecessary to include this region when calibrating $c_{2,.}$. For this reason, we only use the data in the convectively stable zone to calibrate $c_{2,.}$. In an early paper \citep{chan2010overshooting}, we have suggested to use the zero points of the vertical velocity as the proxy of upward overshooting boundary. Following this work, \citet{cai2020upward} have identified three layers in the convectively stable zone: the thermal adjustment layer (mixing both entropy and material), the turbulent dissipation layer (mixing material but not entropy), and the thermal dissipation layer (mixing neither entropy nor material). In the thermal adjustment layer, the thermal structure is adjusted so that the radiative flux is able to balance the negative convective flux \citep{deng2008define}. To capture this phenomena as accurate as possible, we calibrate $c_{2,.}$ with the data only in this thermal adjustment layer. In addition, the diffusive terms include the derivatives of the TOMs. A constant intercept will not affect the derivative. For this reason, we keep the nonzero intercept when fitting the TOMs by the SOMs.

Table \ref{tab:table1} presents the coefficients calibrated from the 3D data. The last two rows give the mean and standard deviation among all the simulated cases for the calibrated coefficients, respectively. The mean values of $c_{1,w^2}$, $c_{1,\theta^2}$, $c_{1,w\theta}$, and $c_{3}$ are 1.15, 4.97, 0.65, and 5.15 respectively. These values do not vary too much across different cases, with dispersions of 7 percent, 26 percent, 5 percent and 2 percent, respectively. These calibrated coefficients do not differ much from those estimated by the data of pure convection zones \citep{cai2018numerical}. Thus, we conclude that the convective and isotropic coefficients could hardly be affected by the adjacent stable zones. The last column also lists the correlation coefficient between the vertical velocity and temperature perturbation. As confirmed in previous simulations \citep{cai2018numerical}, these correlation coefficients are close to 0.6.

\begin{figure}
\gridline{\fig{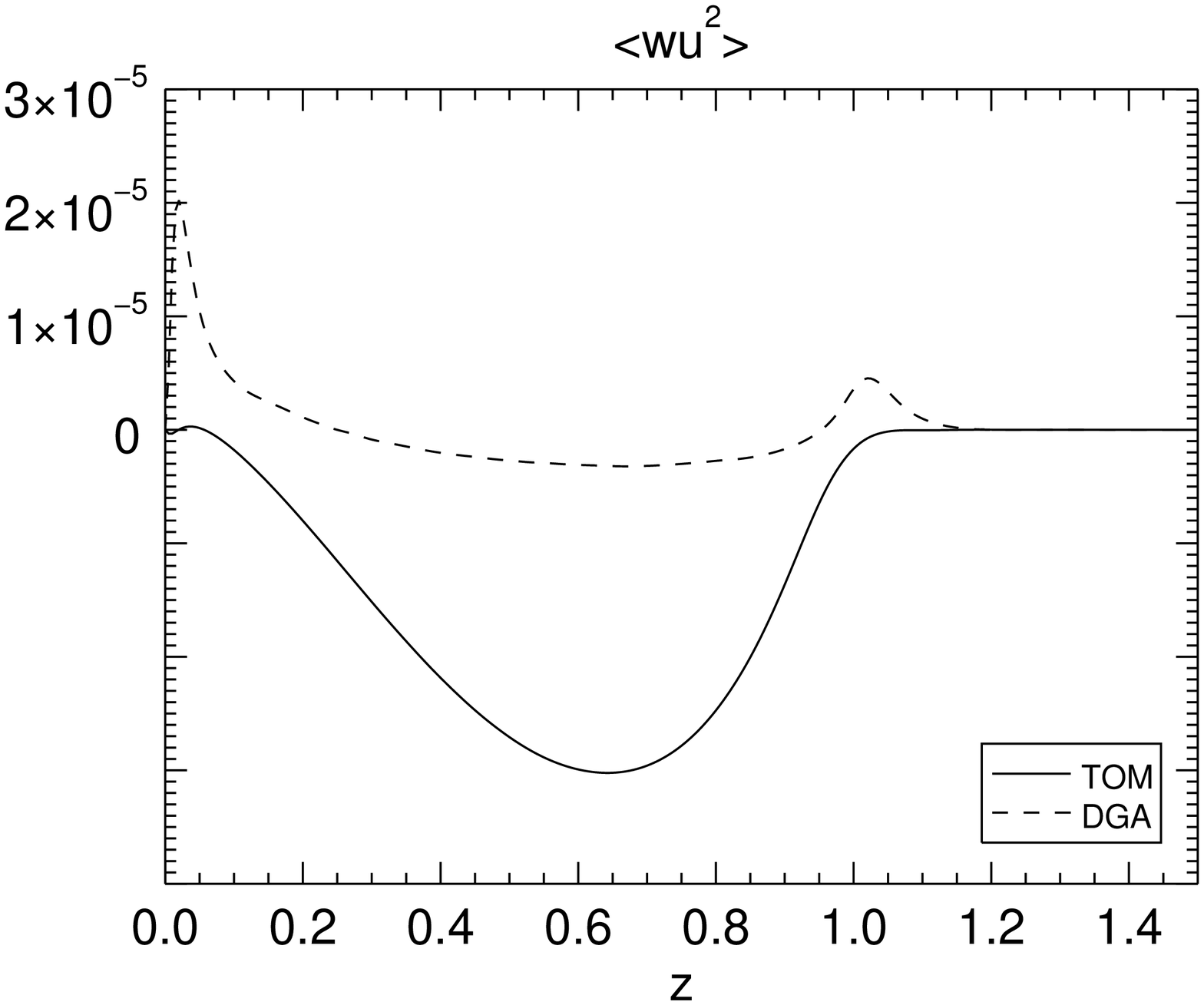}{0.45\textwidth}{(a)}
          \fig{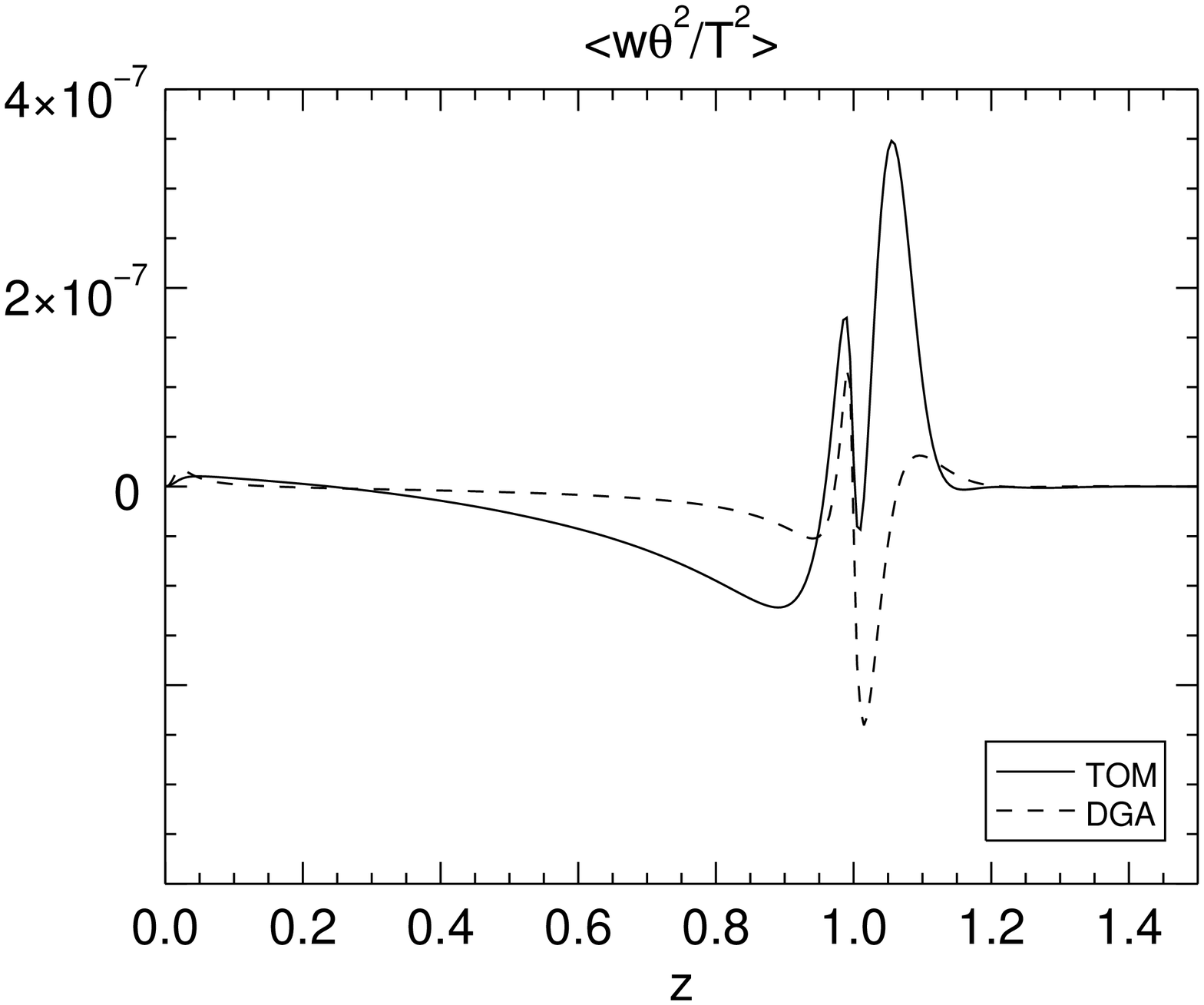}{0.45\textwidth}{(b)}
          }
\gridline{\fig{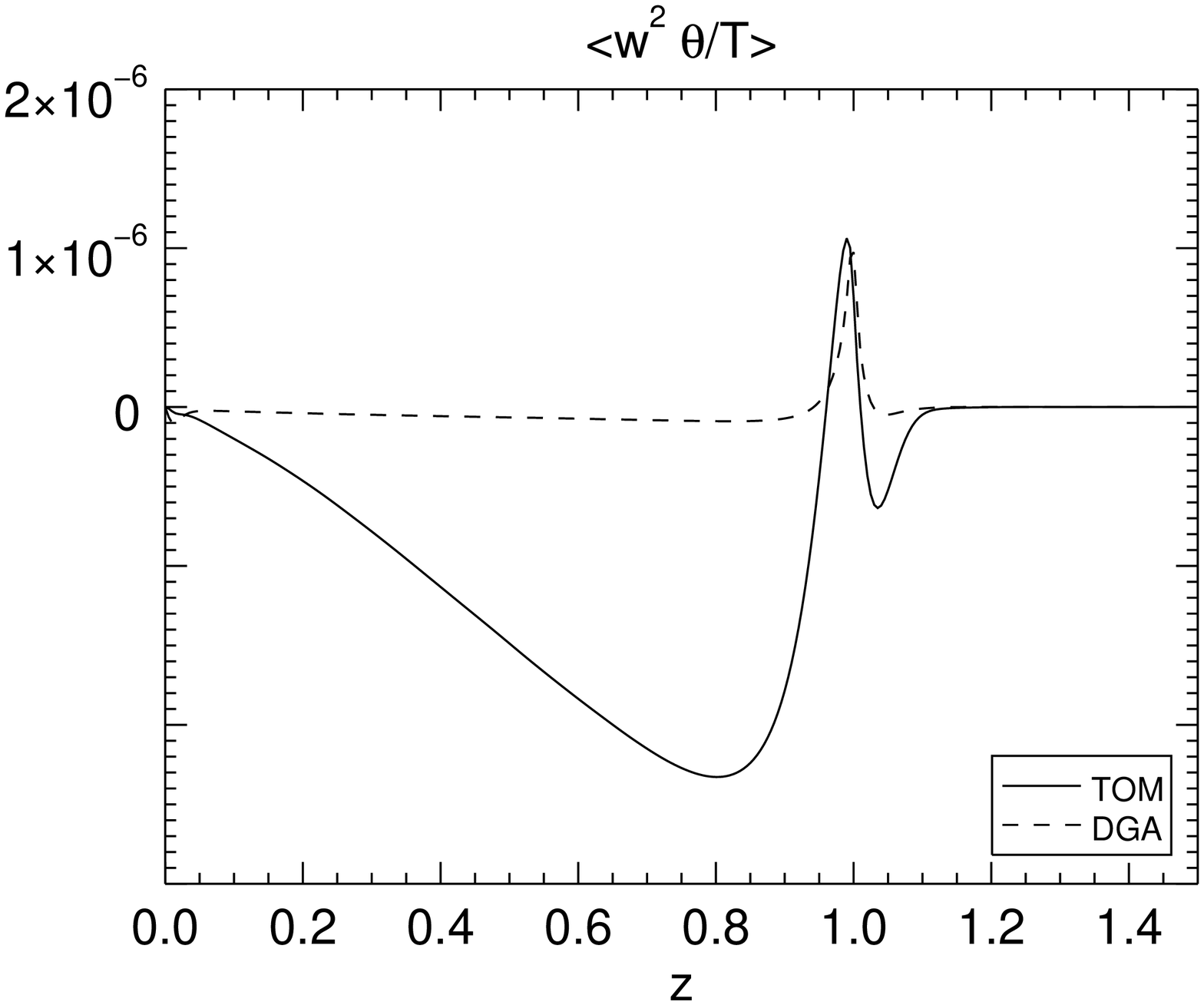}{0.45\textwidth}{(c)}
          \fig{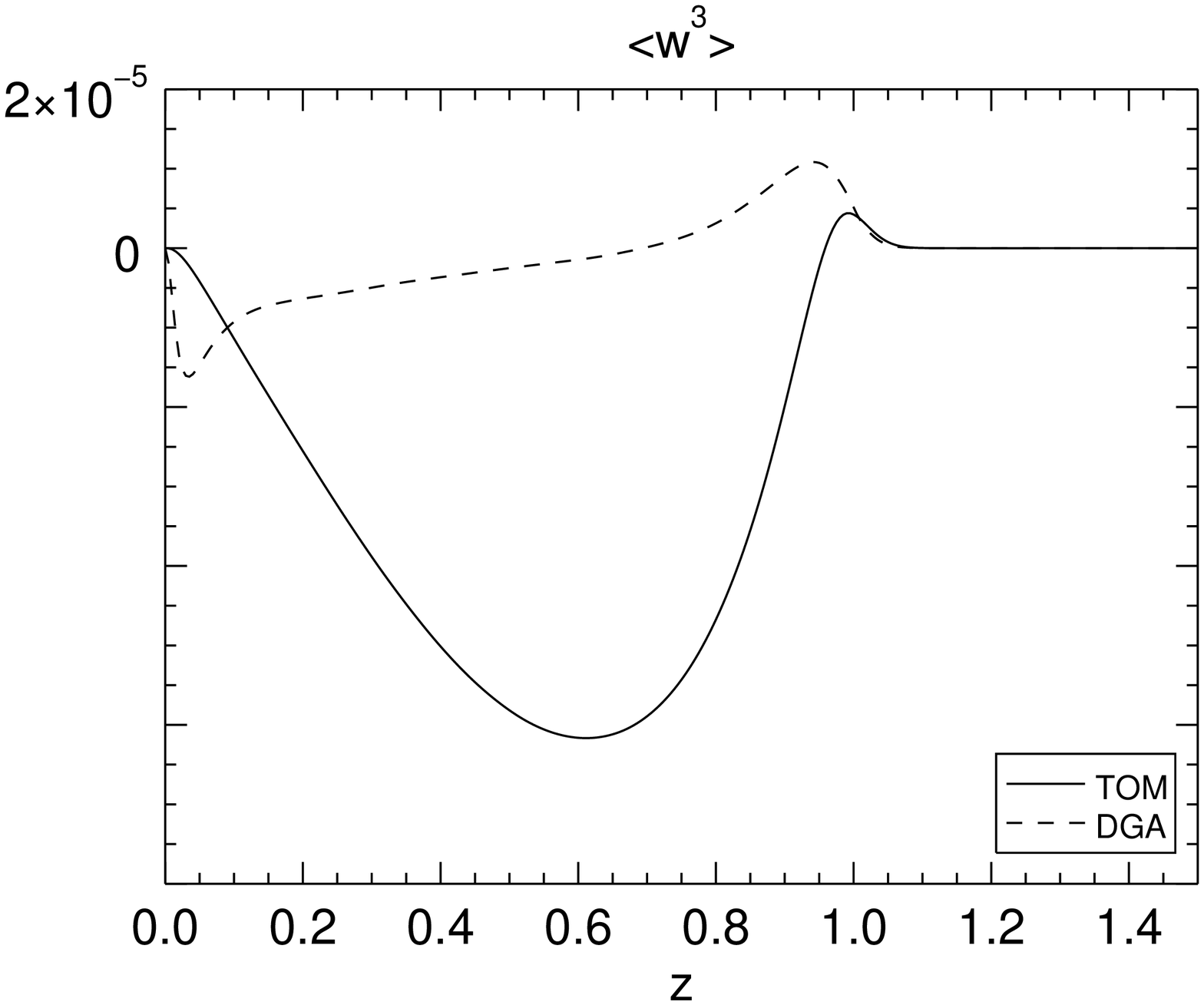}{0.45\textwidth}{(d)}
          }
\caption{The third-order moments and the downgradient approximations of third-order moments calculated from the 3D data of the case A3. Panels (a)-(d) show the TOM (solid line) and DGA (dashed line) of $\langle \overline{wu^2}\rangle$, $\langle \overline{w\theta^2/T^2} \rangle$, $\langle \overline{w^2 \theta/T}\rangle$, and $\langle \overline{w^3}\rangle$, respectively. \label{fig:f2}}
\end{figure}

\begin{figure}
\gridline{\fig{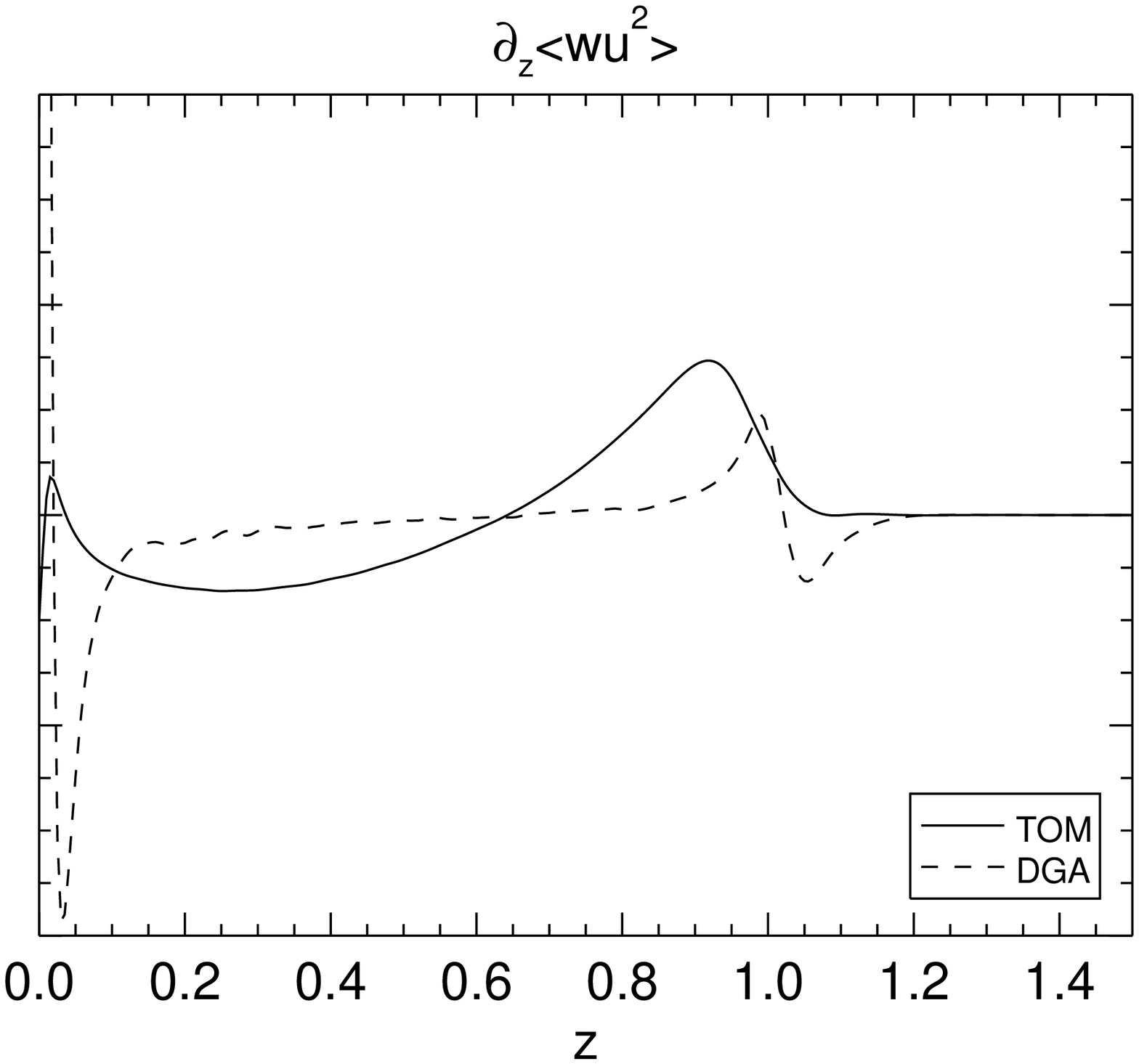}{0.45\textwidth}{(a)}
          \fig{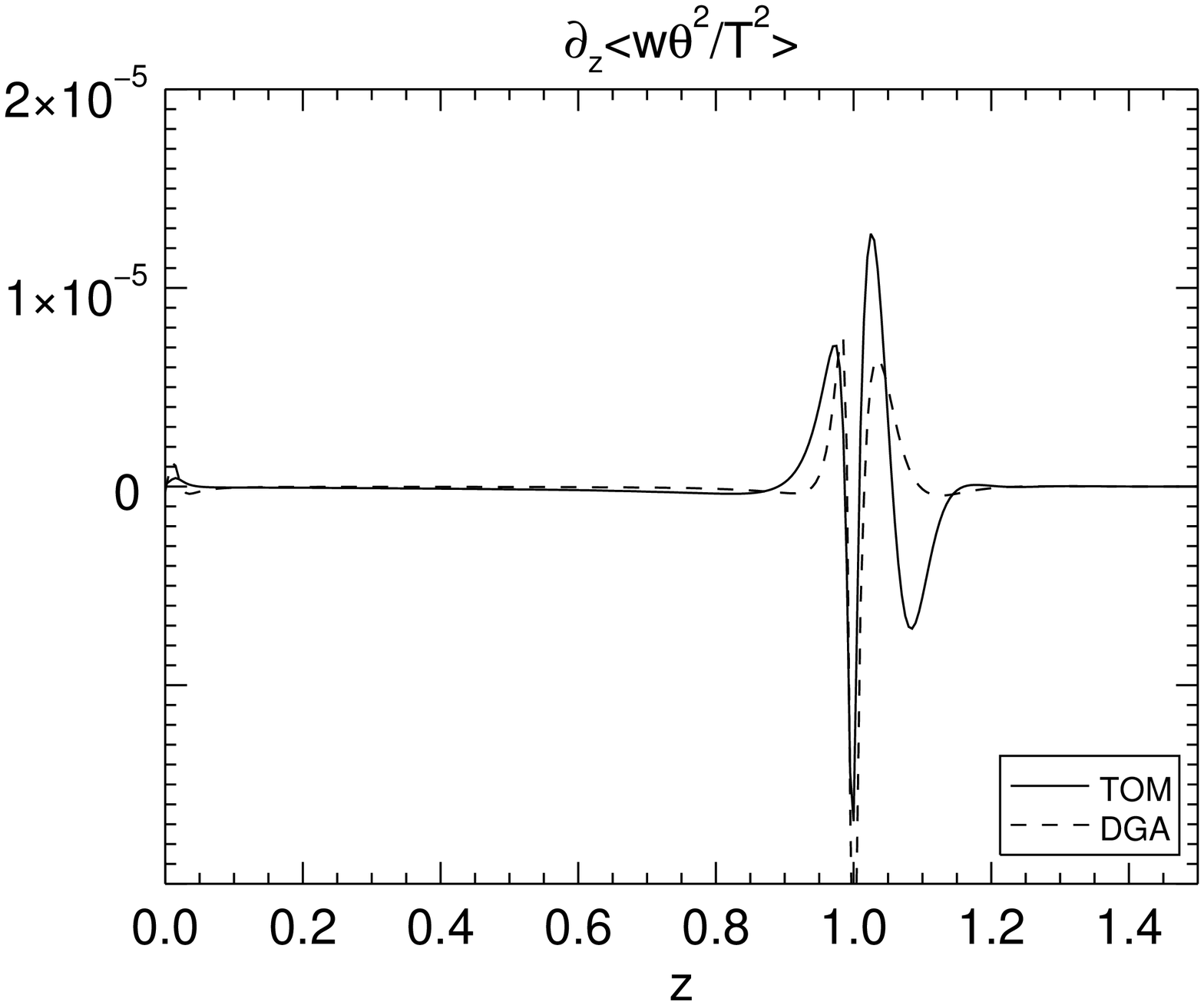}{0.45\textwidth}{(b)}
          }
\gridline{\fig{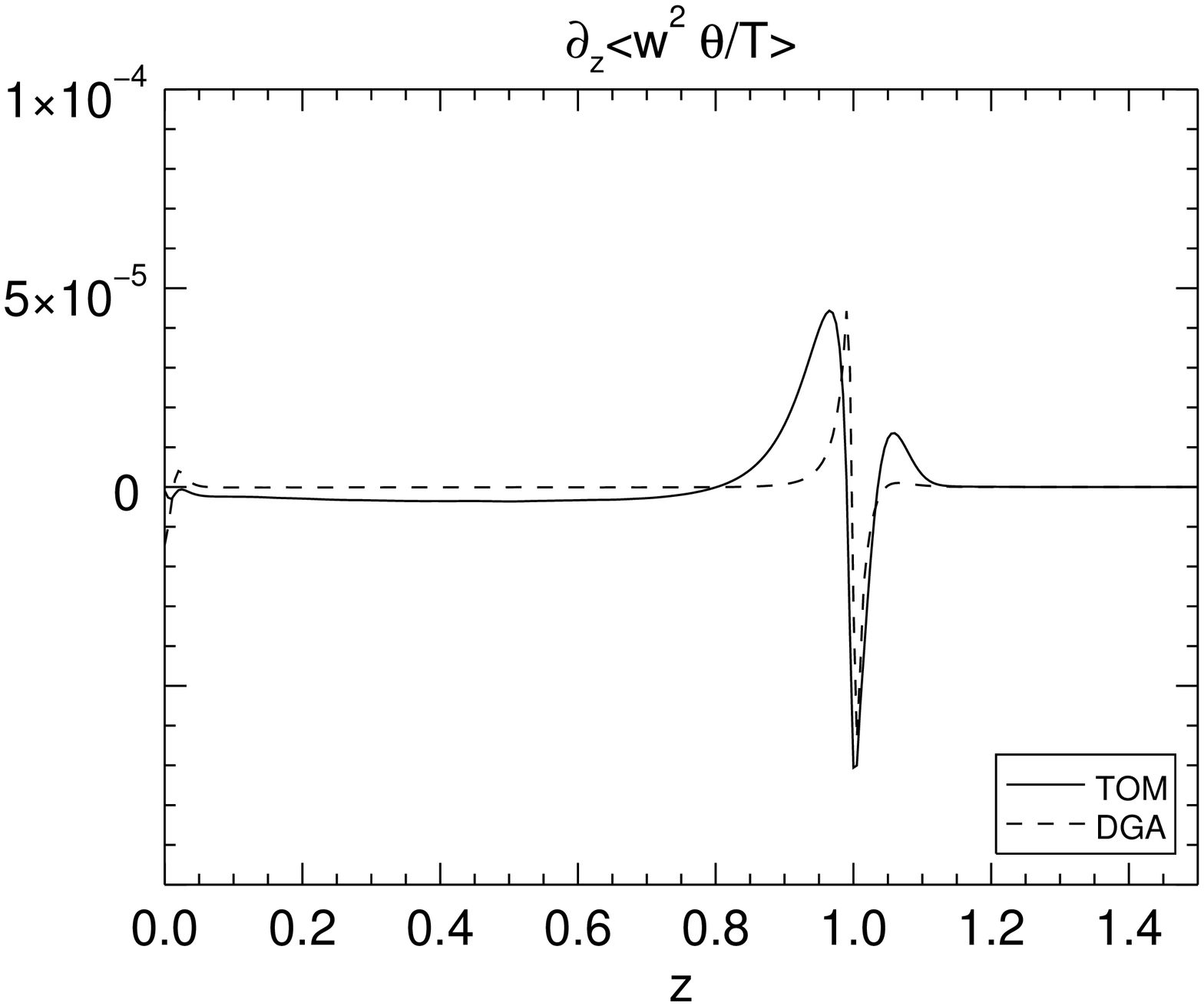}{0.45\textwidth}{(c)}
          \fig{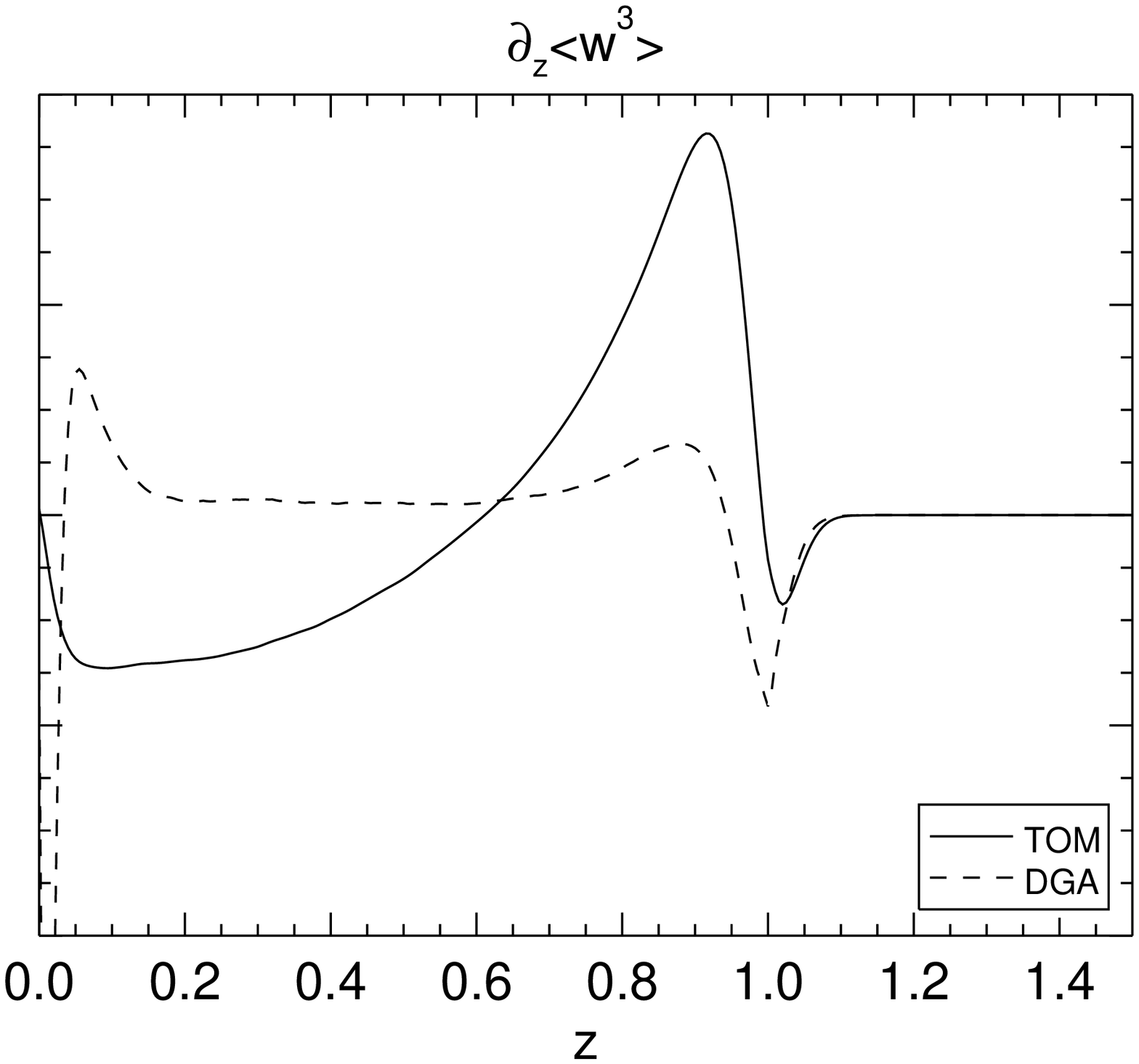}{0.45\textwidth}{(d)}
          }
\caption{The derivatives of the third-order moments and the downgradient approximations of third-order moments calculated from the 3D data of the case A3. Panels (a)-(d) show the TOM (solid line) and DGA (dashed line) of $\partial_{z}\langle \overline{wu^2}\rangle$, $\partial_{z}\langle \overline{w\theta^2/T^2}\rangle$, $\partial_{z}\langle \overline{w^2 \theta/T}\rangle$, and $\partial_{z}\langle \overline{w^3}\rangle$, respectively.\label{fig:f3}}
\end{figure}

The mean values of the calibrated diffusive coefficients $c_{2,w^2}$, $c_{2,\theta^2}$, and $c_{2,w\theta}$ are 0.26, 0.20, and 0.08, respectively. $c_{2,\theta^2}$ is close to the one estimated in \citet{cai2018numerical}. However, $c_{2,w^2}$ and $c_{2,w\theta}$ are much smaller than those calibrated in \citet{cai2018numerical}. The methods for calibrations of the diffusive coefficients are different in these two papers. \citet{cai2018numerical} calibrated $c_{2,.}$ by the diffusive effect induced by the boundary conditions. It seems that the diffusive effect induced by the boundary conditions is stronger than by the adjacent stable zone. Now we check the validity of the DGAs by comparing the left with the right hand sides of eqs.(\ref{eq:wu2}-\ref{eq:w3}). Fig.\ref{fig:f2} depicts the TOMs and the DGAs of these TOMs with the 3D data. Obviously, the performance of the DGAs is unsatisfactory in the convectively unstable zone. However, the DGAs do capture some properties in the overshooting zone. For example, the dips and bumps of $\langle \overline{w\theta^2/T^2} \rangle$, $\langle \overline{w^2 \theta/T}\rangle$, and $\langle \overline{w^3}\rangle$ near the interface are replicated by the DGAs. In addition, the widths of these dips and bumps of DGAs agree well with the TOMs. DGAs correctly predict the signs of the dips and bumps on $\langle \overline{w^2 \theta/T}\rangle$ and $\langle \overline{w^3}\rangle$. However, the prediction on the sign of $\langle \overline{w\theta^2/T^2} \rangle$ is not good. In the 1D RSM, the derivatives of the TOMs are involved in the eqs.(\ref{eq:u2}-\ref{eq:w2}). Thus, the predictions on the derivatives of the TOMs are more important than the TOMs themselves. Fig.\ref{fig:f3} shows the derivatives of TOMs and DGAs. Now we see that the derivatives of DGA can correctly predict the signs of the bumps and dips of $\partial_{z}\langle \overline{w\theta^2/T^2}\rangle$. Although the DGAs differ significantly from the TOMs in the convectively unstable zone, the differences between their derivatives are diminished.

\begin{deluxetable*}{ccccccccccccccccc}[htb!]
\tablecaption{Estimated coefficients of Xiong's nonlocal model \label{tab:table1}}
\tablehead{
 Case & S & $\mu$ & $F_{tot}$ & $\rm Pr$ & $\rm Re$ & $\rm Pe$ & $c_{1,w^2}$ & $c_{1,\theta^2}$ & $c_{1,w\theta}$ & $c_{2,w^2}$ & $c_{2,\theta^2}$ & $c_{2,w\theta}$ & $c_{3}$  & $cor[w,\theta]$
}
\startdata
 A1 & $1$   &   $1.25\times 10^{-4}$    & 0.00125    & $0.5$    & 1102  & 550.8  & 1.22 & 7.33 & 0.64 & 0.36 & 0.26 & 0.20 & 7.30  & 0.59 \\
 A2 & $2$   &   $1.25\times 10^{-4}$    & 0.00125    & $0.5$    & 1084  & 542.2  & 1.10 & 6.57 & 0.63 & 0.28 & 0.18 & 0.13 & 4.64  & 0.60 \\
 A3 & $3$   &   $1.25\times 10^{-4}$    & 0.00125    & $0.5$    & 1062  & 530.9  & 1.16 & 4.76 & 0.62 & 0.26 & 0.14 & 0.09 & 5.60  & 0.60 \\
 A4 & $4$   &   $1.25\times 10^{-4}$    & 0.00125    & $0.5$    & 1062  & 530.8  & 1.13 & 4.86 & 0.64 & 0.25 & 0.13 & 0.07 & 4.81  & 0.61 \\
 A5 & $5$   &   $1.25\times 10^{-4}$    & 0.00125    & $0.5$    & 1059  & 529.6  & 1.10 & 4.50 & 0.62 & 0.24 & 0.10 & 0.04 & 4.27  & 0.61 \\
 A6 & $6$   &   $1.25\times 10^{-4}$    & 0.00125    & $0.5$    & 1055  & 527.2  & 1.15 & 4.13 & 0.70 & 0.24 & 0.07 & 0.02 & 4.72  & 0.64 \\
 A7 & $7$   &   $1.25\times 10^{-4}$    & 0.00125    & $0.5$    & 1034  & 516.9  & 1.14 & 2.87 & 0.70 & 0.27 & 0.08 & 0.02 & 5.24  & 0.67 \\
 B1 & $3$   &   $2.5\times 10^{-4}$     & 0.00125    & $1.0$    & 509   & 509.0  & 0.99 & 5.94 & 0.61 & 0.30 & 0.12 & 0.09 & 3.98  & 0.62 \\
 B2 & $3$   &   $6.25\times 10^{-5}$    & 0.00125    & $0.25$   & 2213  & 553.3  & 1.23 & 4.75 & 0.68 & 0.24 & 0.13 & 0.09 & 5.53  & 0.61 \\
 B3 & $3$   &   $3.125\times 10^{-4}$   & 0.00125    & $0.125$  & 4489  & 561.1  & 1.28 & 4.32 & 0.68 & 0.22 & 0.14 & 0.09 & 5.80  & 0.61 \\
 C1 & $3$   &   $2.5\times 10^{-4}$     & 0.00250    & $0.5$    & 669   & 334.5  & 1.07 & 6.52 & 0.62 & 0.36 & 0.19 & 0.13 & 4.39  & 0.61 \\
 C2 & $3$   &   $6.25\times 10^{-5}$    & 0.000625   & $0.5$    & 1704  & 851.8  & 1.19 & 4.26 & 0.66 & 0.20 & 0.11 & 0.07 & 5.02  & 0.62 \\
 C3 & $3$   &   $3.125\times 10^{-5}$   & 0.0003125  & $0.5$    & 2730  & 1364.2 & 1.23 & 3.76 & 0.66 & 0.15 & 0.07 & 0.05 & 5.66  & 0.61 \\
 \hline
 avg & - & - & - & - & - & - & 1.15 & 4.97 & 0.65 & 0.26 & 0.20 & 0.08 & 5.15 & 0.62\\
 dev & - & - & - & - & - & - & 0.08 & 1.27 & 0.03 & 0.06 & 0.09 & 0.05 & 0.86 & 0.02\\
\enddata
\tablecomments{$S$ is the stability parameter. $\mu$ is the dynamic viscosity. $F_{tot}$ is the total flux. $\rm Pr$ is the Prandtl number. $\rm Re$ is the averaged Reynolds number. $\rm Pe$ is the averaged P\'eclet number. The averages are taken both temporarily and spatially in the convectively unstable zone. $c_{1,w^2},c_{1,\theta^2},c_{1,w\theta}$ are the convective coefficients. $c_{2,w^2},c_{2,\theta^2},c_{2,w\theta}$ are the diffusive coefficients. $c_{3}$ is the coefficient measuring the isotropic level of fluid motions.  $cor[w,\theta]=\overline{w\frac{\theta}{T}}/(\overline{w^2}\overline{\frac{\theta^2}{T^2}})^{1/2}$ represents the correlation coefficient between vertical velocity and temperature perturbation. The last two rows give the averages and standard deviations of the estimated coefficients.}
\end{deluxetable*}

\subsection{Comparison between 1D and 3D results}
The full set of 1D RSM include the equations of the thermal structures. As mentioned in the \citet{cai2018numerical}, the mass, momentum, and energy conservation equations in the 1D RSM are
\begin{eqnarray}
\frac{\partial m}{\partial z}=\rho~,\\
\frac{\partial (P+\rho \overline{u^2})}{\partial z} = -\rho g~, \label{eq:balance}\\
F_{r}+F_{c}+F_{k}=F_{tot}~,
\end{eqnarray}
where $m$ is the mass; $g$ is the gravitational acceleration; $F_{r}=-\kappa_{T} \frac{\partial T}{\partial z}$ is the conductive flux (or radiative flux), $F_{c}=\rho c_{p}T \overline{w\frac{\theta}{T}}$ is the convective flux; $F_{k}=-\frac{3\sqrt{3}}{8}c_{2,w^2}\frac{P}{g} (\overline{w^2})^{1/2} \frac{\partial}{\partial z}\overline{u^2}$ is the kinetic energy flux; and $F_{tot}$ is the total flux. Given the coefficients and appropriate boundary conditions, we can solve the full set of 1D RSM equations. We use the same boundary conditions as those applied in 3D simulations \citep{cai2018numerical,cai2020upward}, and we use the calibrated coefficients in table~\ref{tab:table1}.

\begin{figure}
\gridline{\fig{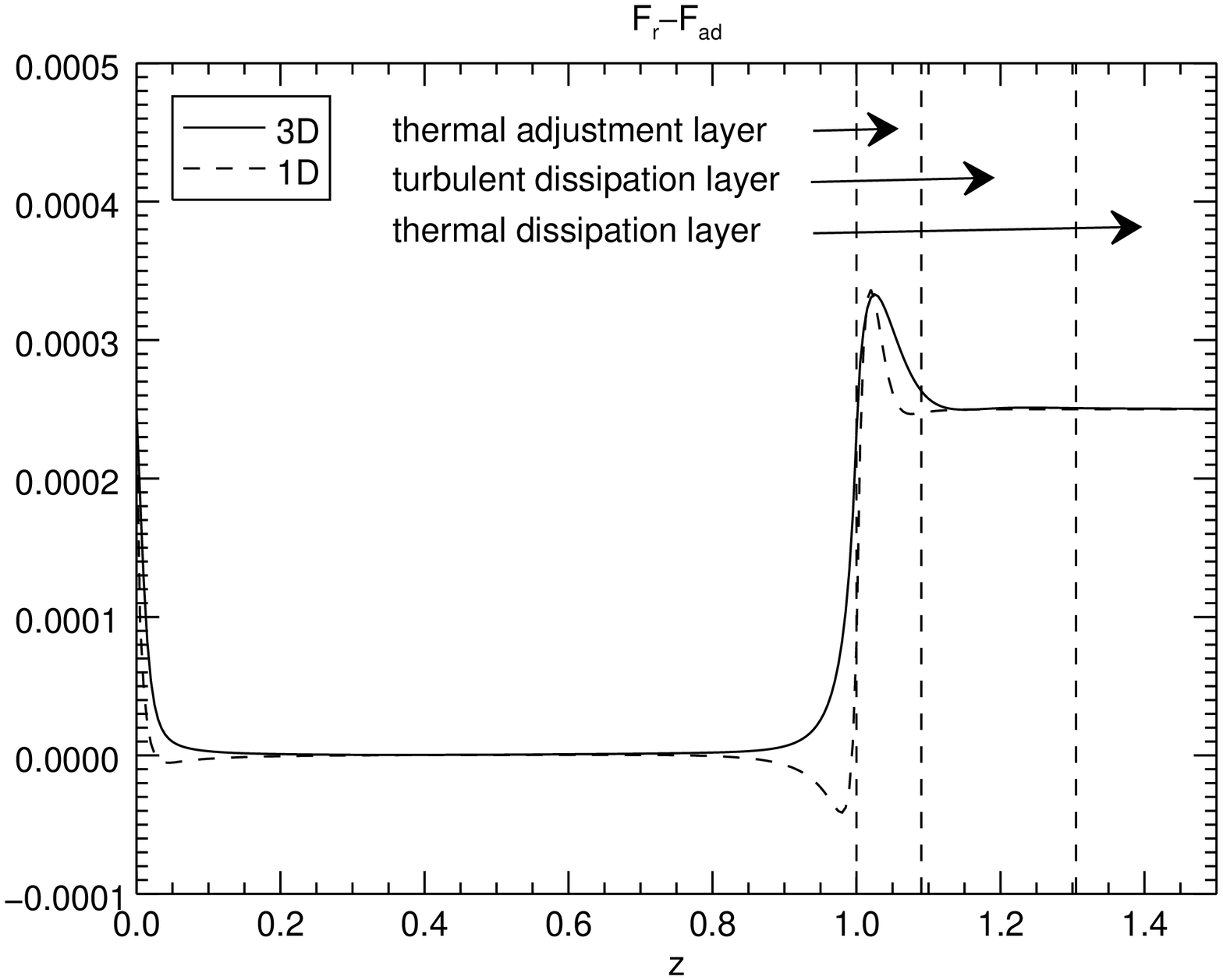}{0.45\textwidth}{(a)}
          \fig{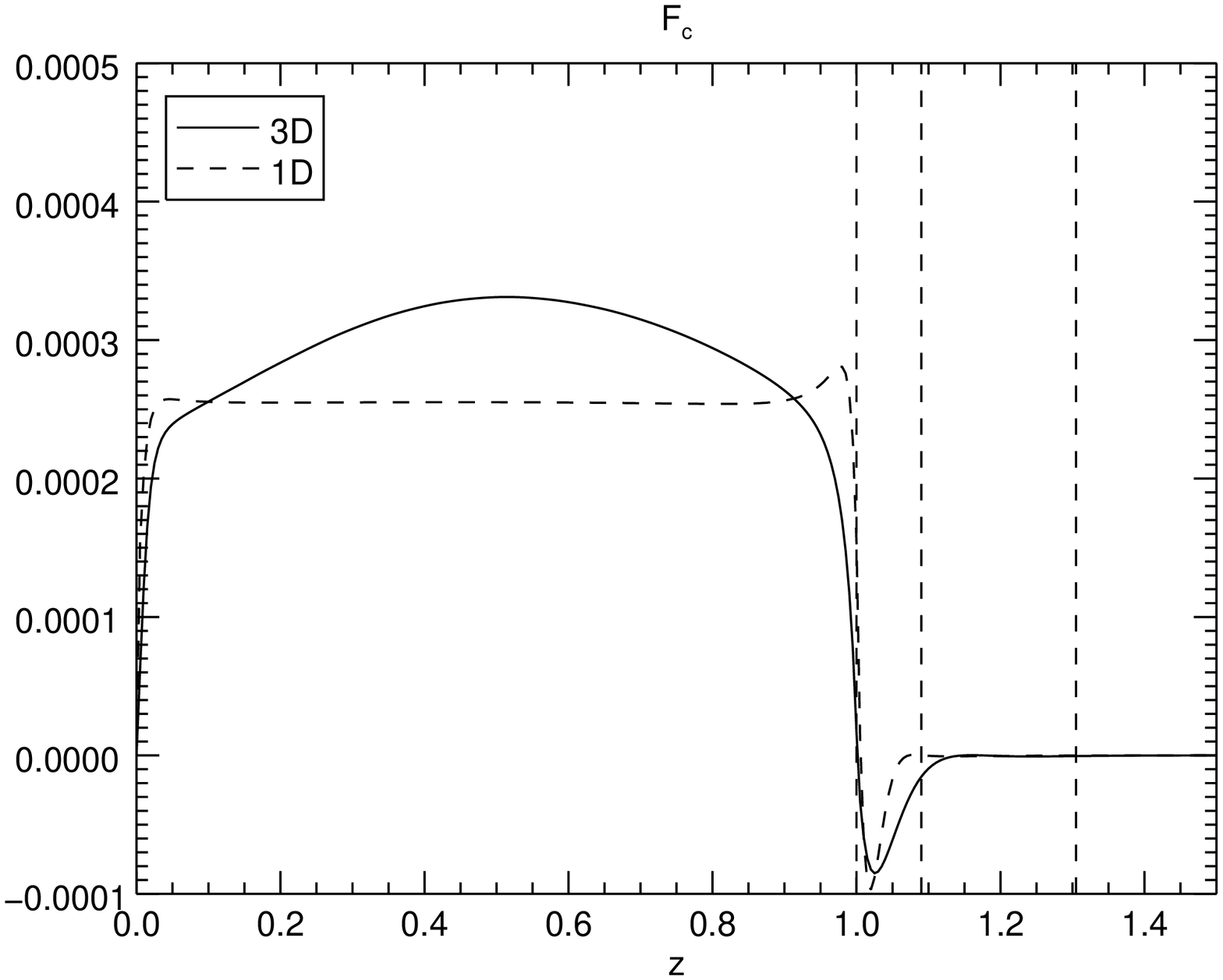}{0.45\textwidth}{(b)}
          }
\gridline{\fig{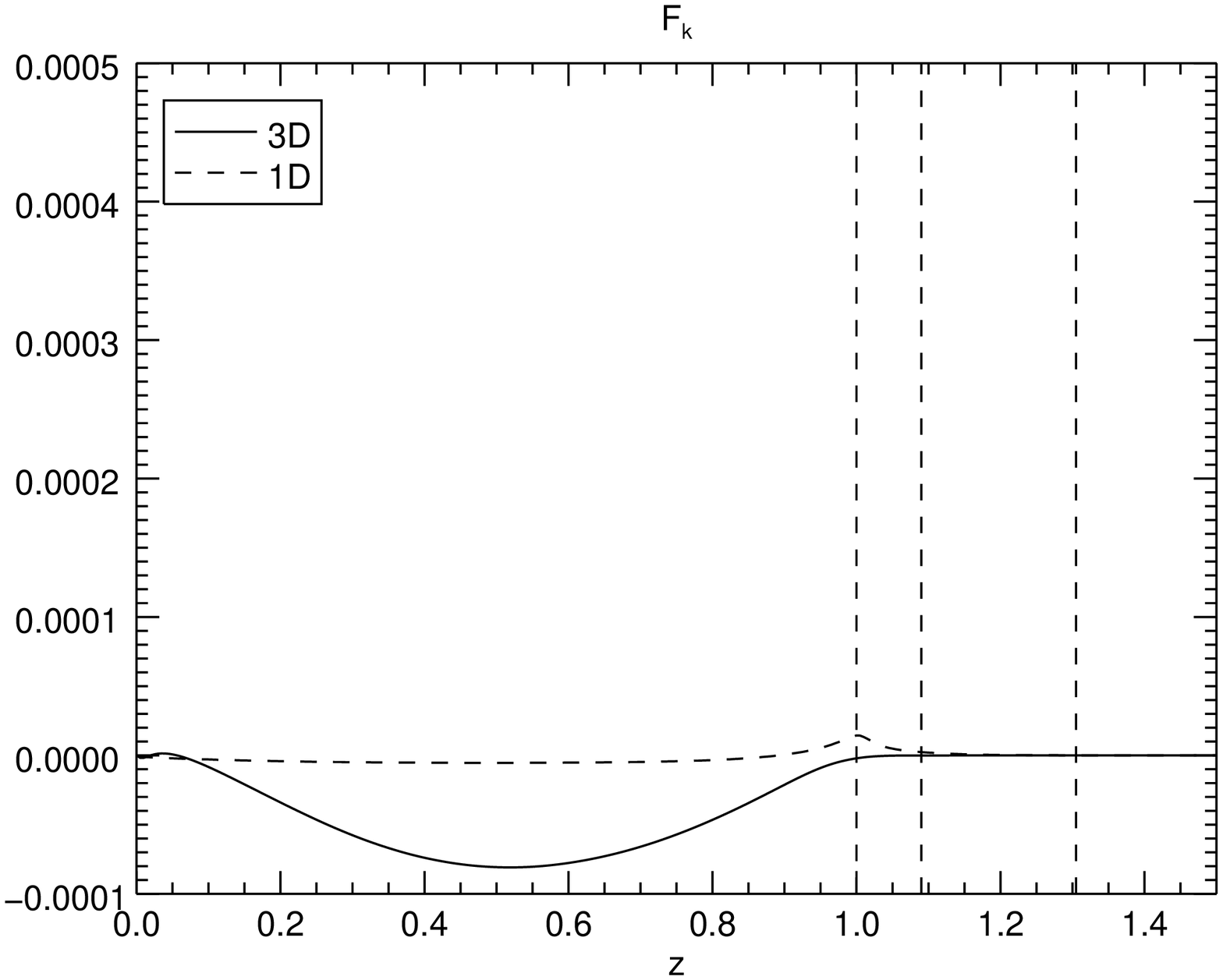}{0.45\textwidth}{(c)}
          \fig{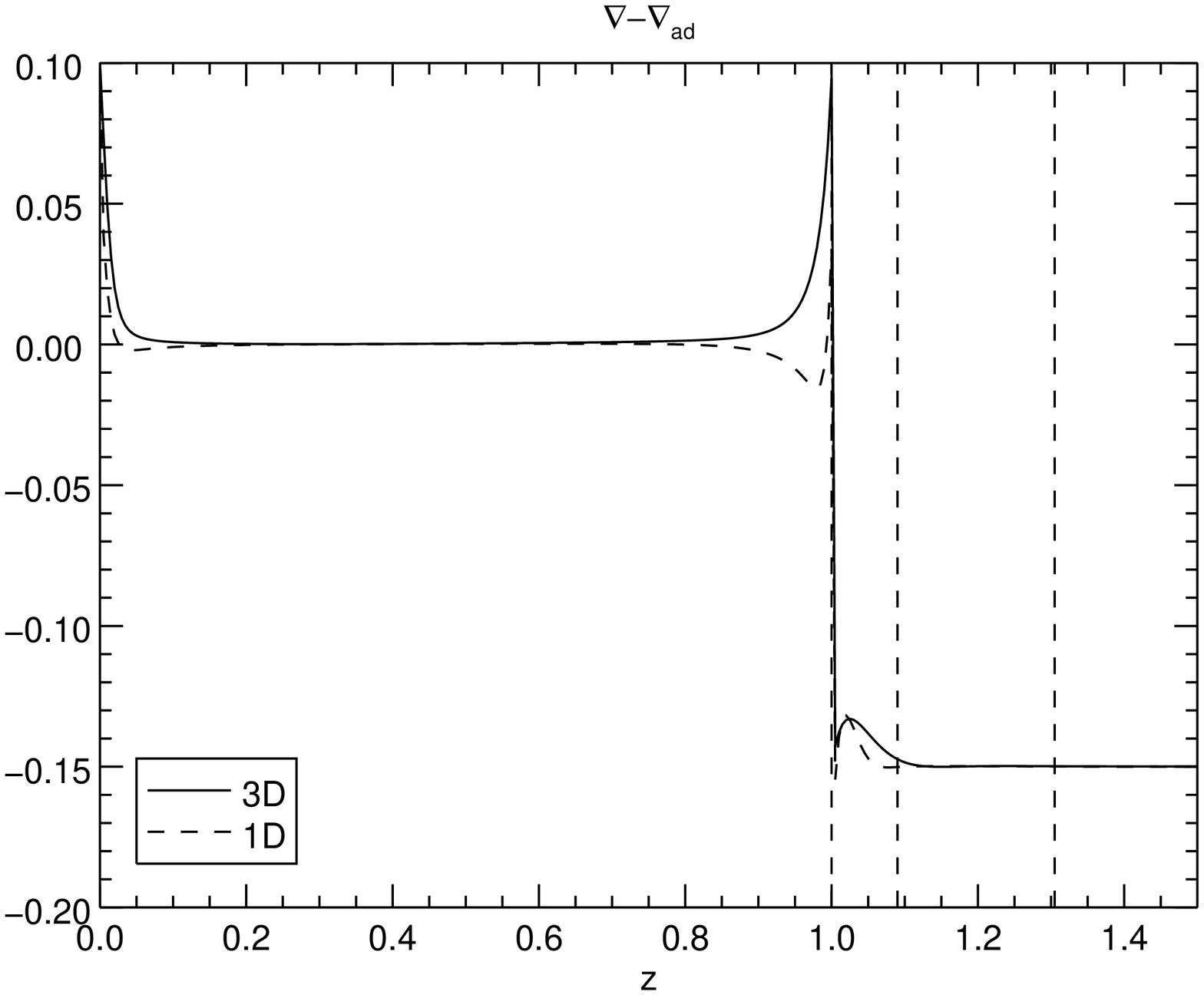}{0.45\textwidth}{(d)}
          }
\caption{Comparison between 1D and 3D results. Panels (a)-(c) present the fluxes $F_{r}-F_{ad}$, $F_{c}$, and $F_{k}$. Panel (d) present the super-adiabatic temperature gradient $\nabla-\nabla_{ad}$. The vertical dashed lines (z=1.0,1.09,1.305) are the locations of the boundaries of thermal adjustment layer, turbulent dissipation layer, and thermal dissipation layer. The shown case is A3. The vertical dashed lines shows the boundaries at the convectively stable/unstable zones, the first zero point of vertical velocity correlation, and the second zero point of vertical velocity correlation.\label{fig:f4}}
\end{figure}

\begin{figure}
\gridline{\fig{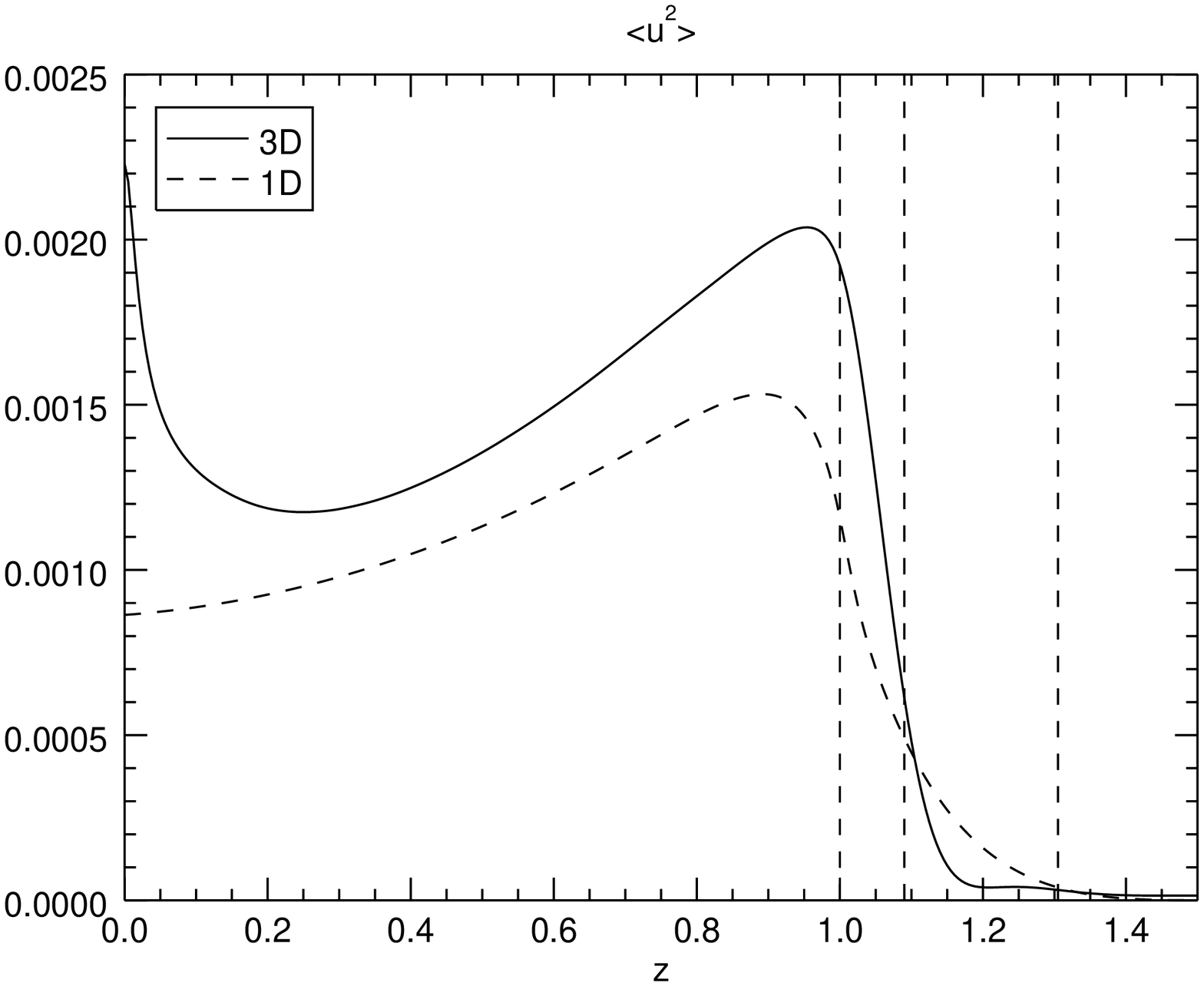}{0.45\textwidth}{(a)}
          \fig{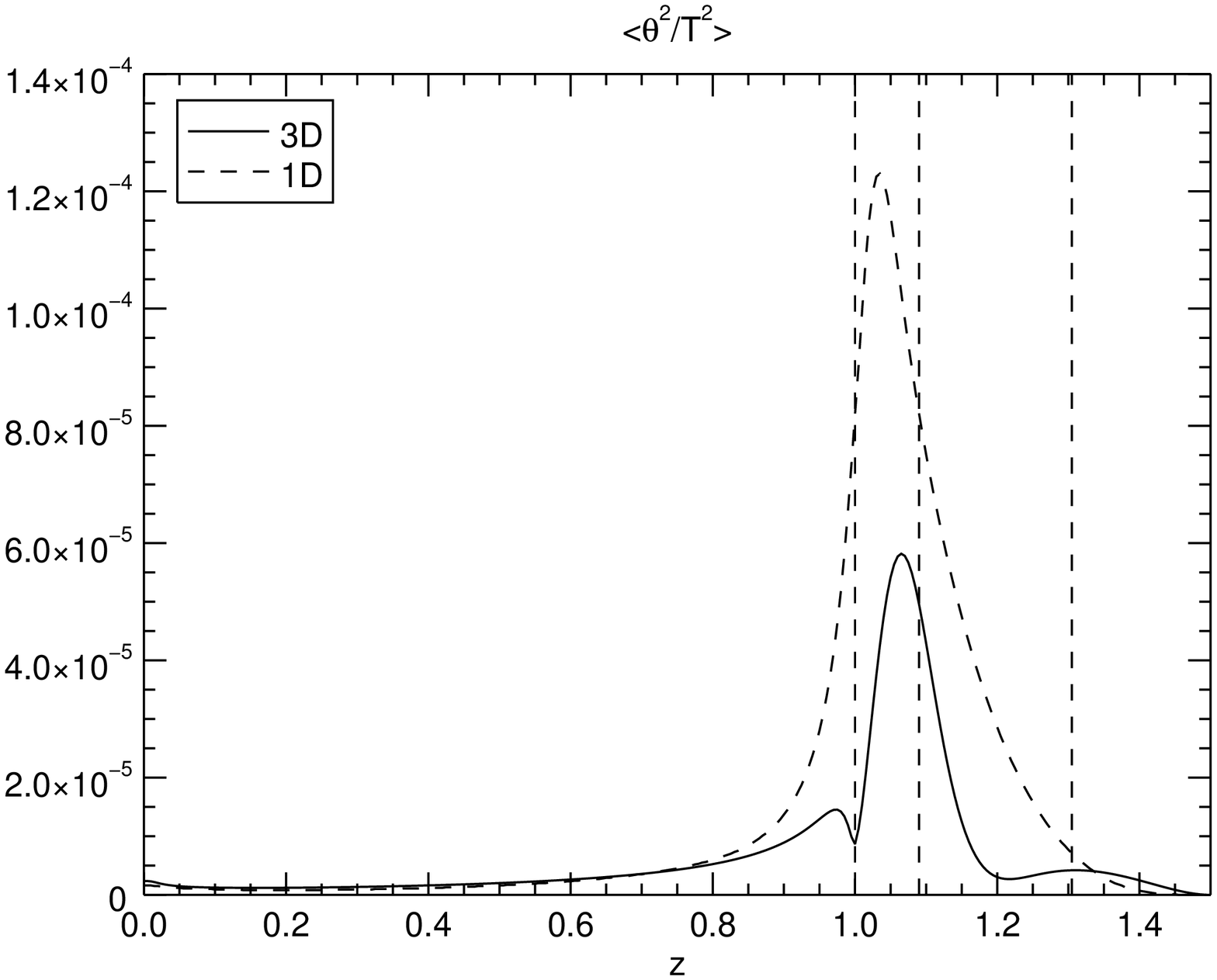}{0.45\textwidth}{(b)}
          }
\gridline{\fig{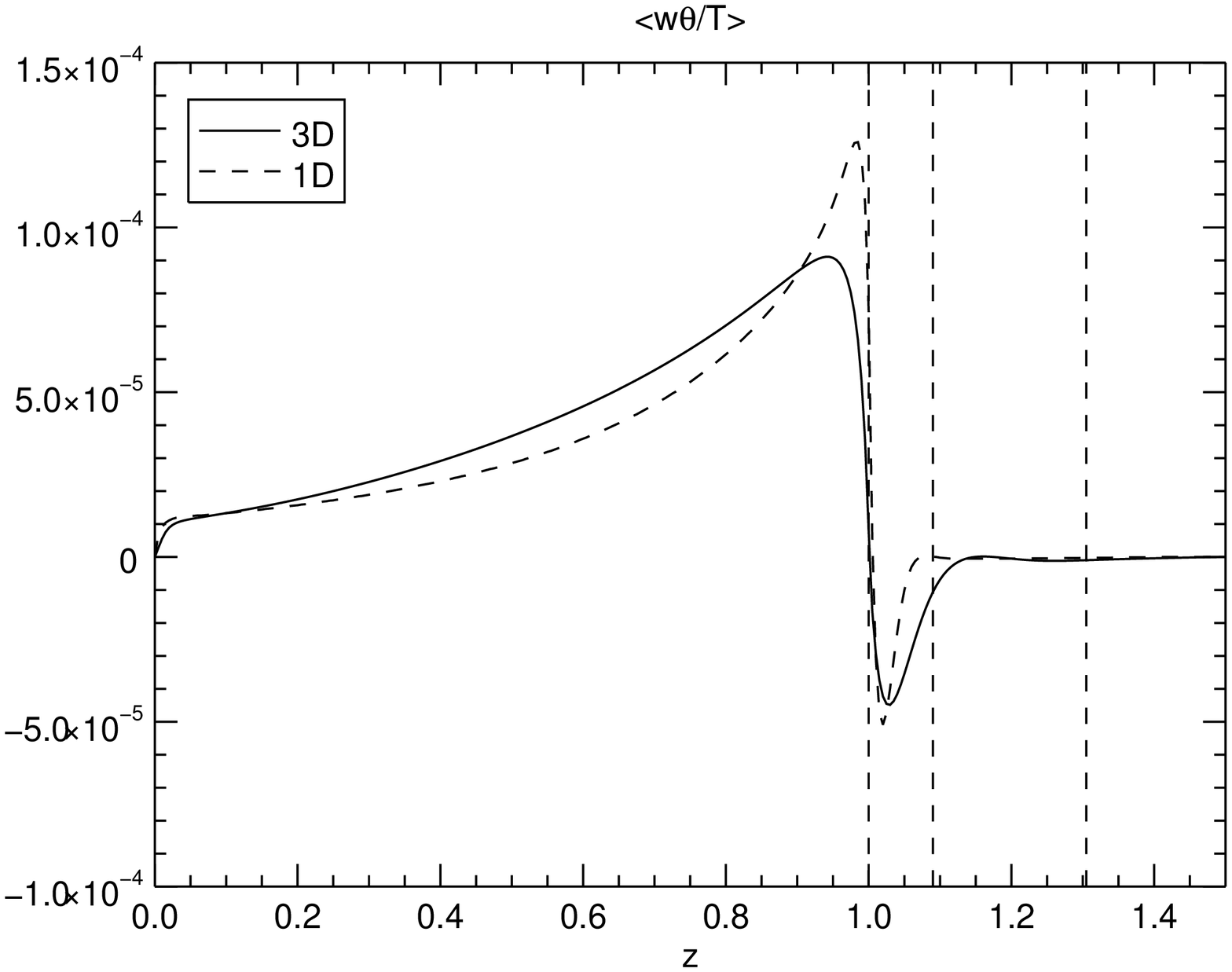}{0.45\textwidth}{(c)}
          \fig{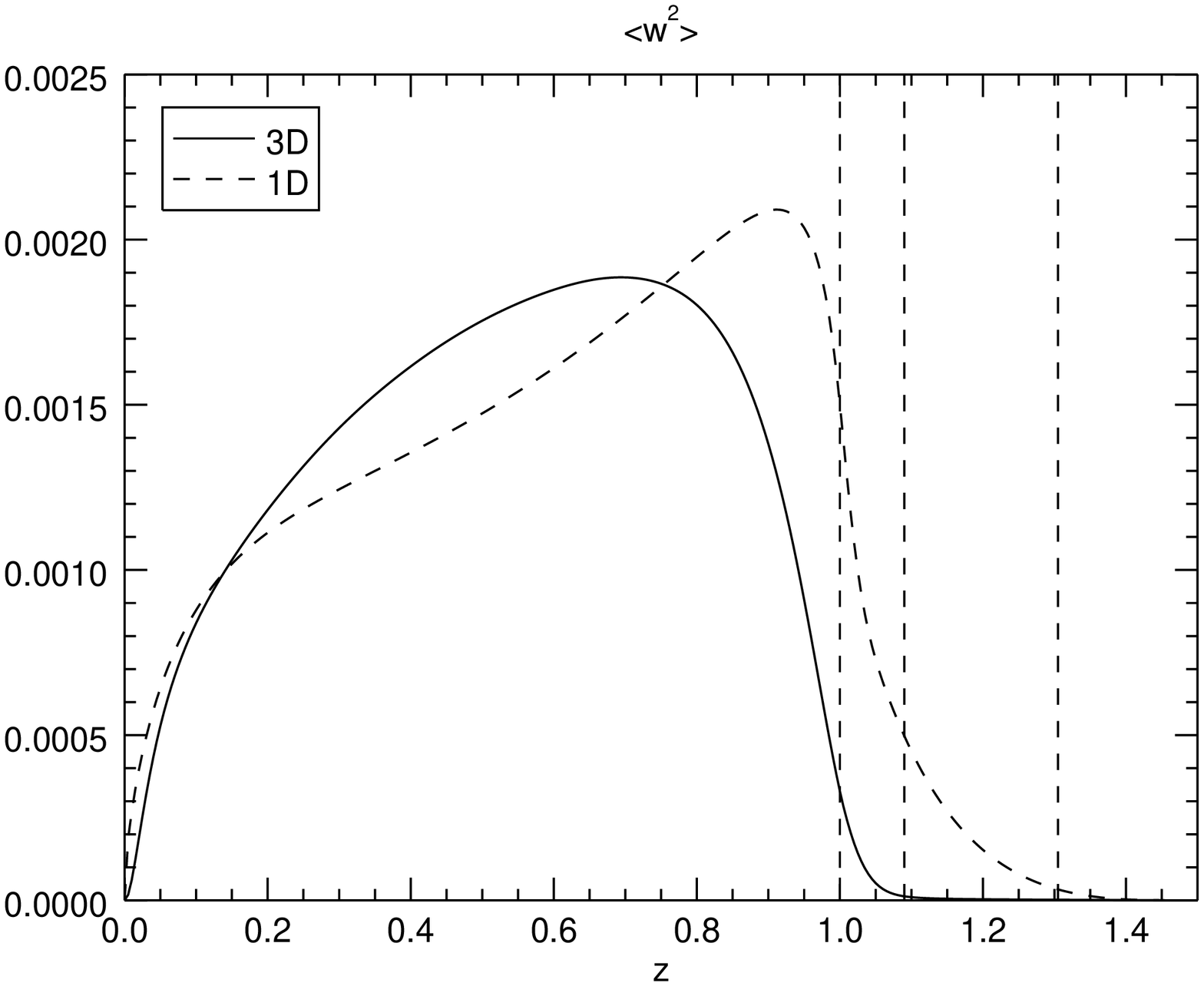}{0.45\textwidth}{(d)}
          }
\gridline{\fig{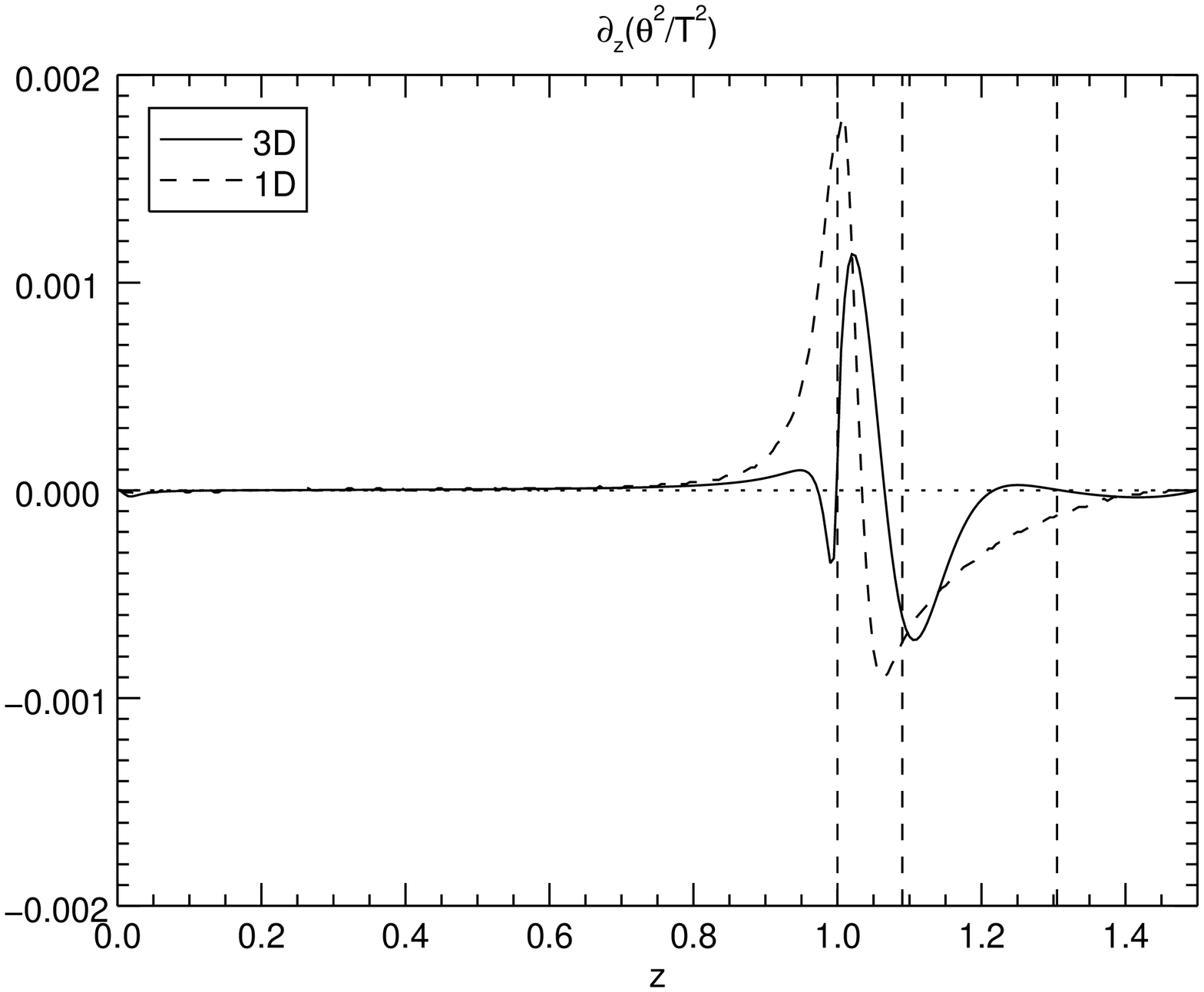}{0.45\textwidth}{(e)}
          \fig{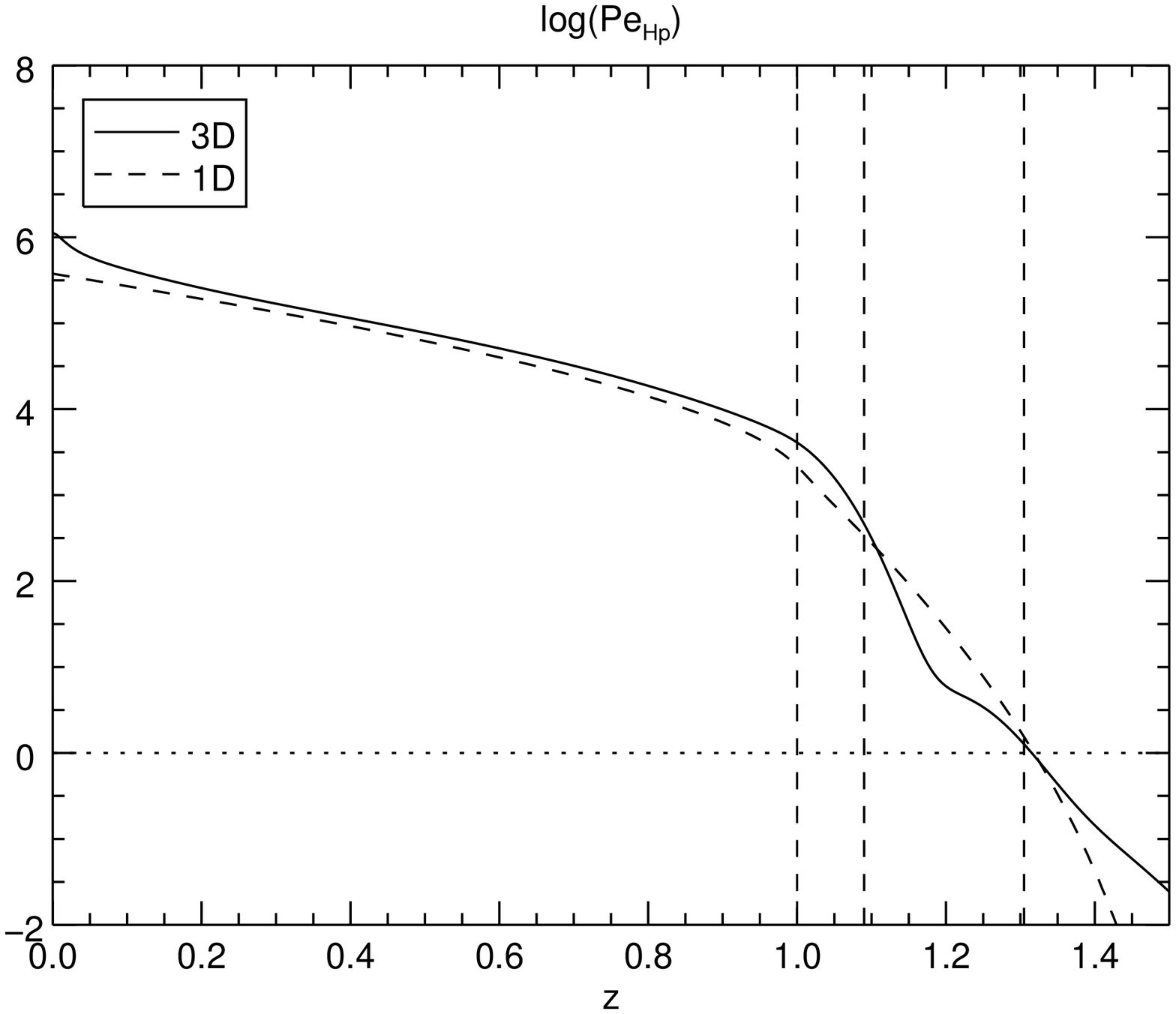}{0.45\textwidth}{(f)}
          }
\caption{Comparison between 1D and 3D results. Panels (a)-(d) presents the profiles of the second-order moments $\langle\overline{u^2}\rangle$, $\langle\overline{\theta^2}\rangle$, $\langle\overline{w\theta/T}\rangle$, and $\langle\overline{w^2}\rangle$. Panel (e) presents the profile of $\langle\partial_{z}\overline{\theta^2/T^2}\rangle$. Panel (f) presents the logarithm of P\'eclet number. The vertical dashed lines (z=1.0,1.09,1.305) are the locations of the boundaries of thermal adjustment layer, turbulent dissipation layer, and thermal dissipation layer. The shown case is A3. The vertical dashed lines shows the boundaries at the convectively stable/unstable zones, the first zero point of vertical velocity correlation, and the second zero point of vertical velocity correlation.\label{fig:f5}}
\end{figure}

Fig.\ref{fig:f4} compares the fluxes between 1D and 3D results for the case A3. Panel (a) presents the net conductive flux $F_{r}-F_{ad}$, where the flux transported by adiabatic temperature gradient $F_{ad}=[(m_{1}+1)/(m_{ad}+1)]F_{tot}$ is deducted from the conductive flux. Both the 1D and 3D results show a bump in the thermal adjustment layer. The size and amplitude of the bump are well predicted by the 1D model. In this region, both the material and entropy are mixed. The temperature perturbation switches sign as the upward drafts cross the interface, leading to an anti-correlation between the vertical velocity and temperature perturbation. As a consequence, the convective flux turns to be negative (see fig.\ref{fig:f4}(b)). To balance the negative convective flux, the temperature gradient has to increase to be super-adiabatic $\nabla>\nabla_{ad}$ (see fig.\ref{fig:f4}(d)). The entropy can hardly be mixed above the thermal adjustment layer, thus almost all the energy is transported by conduction over there. Apart from this similarity, the 1D result differs from the 3D result in several aspects. First, the convective flux of the 1D result is almost equal to $F_{tot}-F_{ad}$ in the convectively unstable zone. However, that of the 3D result can exceed $F_{tot}-F_{ad}$ by about 30 percent. Second, the kinetic energy flux $F_{k}$ of the 1D result is negligible in the convectively unstable zone. By contrast, $F_{k}$ of the 3D result is negative and its magnitude is comparable to that of the convective flux in this region. In the stratified convection zones, the cold concentrated fast downward flows overcome the hot broad slow flows in a horizontal averaged sense, hence resulting in a downward $F_{k}$. As a consequence, the enthalpy flux must excess the total flux to achieve the energy balance. Third, the temperature gradient $\nabla$ of the 1D result is subadiabatic at the top of convection zone. The subadiabatic temperature gradient is also reported in \citet{cai2014numerical}, where a simplified version of 1D RSM is solved. On the contrary, $\nabla_{ad}$ of the 3D result is positive over there. This difference can be explained by looking at the SOMs. Fig.\ref{fig:f5}(d) clearly shows that the vertical velocity of the 3D case starts to decrease far before reaching the interface. However, the vertical velocity of the 1D case still remains large close to the interface. Also note that the temperature perturbation of the 1D case increases dramatically near the interface. As a result, more energy flux is carried by the convection, and $\nabla$ can be subadiabatic.

Figs.\ref{fig:f5}(a)-(d) compare the SOMs between the 1D and 3D results. Obviously, the 1D RSM predicts the SOMs much better than the TOMs. The bumps and dips of SOMs shown in the 3D data are well captured in the 1D RSM. The major difference is that turbulent flows of the 3D case can 'feel' the stability effect much further away from the interface. Thus the entrance velocity (at the interface) of the 3D case is much smaller. As the entrance velocity is much larger for the 1D case, it would be expected that the extent of the overshooting distance is further in the 1D case. However, from the fig.\ref{fig:f5}(b), we see that the temperature perturbation of the 1D case is significantly larger than that of the 3D case. Although the entrance velocity is larger, the braking effect is also stronger. As a result, the width of the thermal adjustment layer does not deviate too much between the 1D and 3D cases. One important question of upward overshooting is on how to determine the extent of overshooting distance. In our previous paper \citep{cai2020upward}, we use the first and second zeros of vertical velocity correlations (with the vertical velocity at the interface) as the proxies. We found that the first zero point is the upper boundary of thermal adjustment layer, and the second zero point is the upper boundary of turbulent dissipation layer. Theoretical analysis on 1D RSM \citep{zhang2012turbulent} suggested to use the peak of $\langle\overline{\theta^2}\rangle$ as the boundary of thermal adjustment layer. However, it seems that this peak in the 1D result is closer to the interface, than the first zero point (see fig.\ref{fig:f5}(b) and (e)). In the 3D simulation of upward overshooting \citep{cai2020upward2}, we have also found that this peak is closer to the interface when the stability parameter is large. Thus the peak of $\langle\overline{\theta^2}\rangle$ is not a good indicator on the boundary of the thermal adjustment layer. From fig.\ref{fig:f5}(e), we see that the inflection point ($\partial_{z^2}\langle\overline{\theta^2/T^2}\rangle=0$ or $\partial_{z^2}\langle\overline{\theta^2}\rangle=0$) is a good candidate as the indictor. This location is closer to the first zero point, in both the 1D and 3D cases. In the analysis of \citet{zhang2012turbulent}, they made an assumption that the diffusion term of $\langle\overline{\theta^2}\rangle$ is ignored. The peak value of $\langle\overline{\theta^2}\rangle$ only guarantees $\partial_{z}\langle\overline{\theta^2}\rangle=0$, whereas the diffusion term of $\langle\overline{\theta^2}\rangle$ still plays a role. On the other hand, the inflection point guarantees $\partial_{z^2}\langle\overline{\theta^2}\rangle=0$, making the assumption more valid around this point.

For the boundary of the turbulent dissipation layer, \citet{zhang2012turbulent} suggested to use the P\'eclet number ${\rm Pe}_{Hp}=\rho c_{p} v'' H_{p}/\kappa_{2}$ as the indictor (${\rm Pe}_{Hp}=1$), where $H_{p}=-\partial z/\partial \log P$ is the pressure scale height. In the fig.\ref{fig:f5}(f), we see that this location agrees well with the second zero point, for both the 1D and 3D cases. Thus we confirm that the location of ${\rm Pe}_{Hp}=1$ is a good indictor on the extent of the turbulent dissipation layer.

\section{Summary}
In a previous paper \citet{cai2020upward}, we have performed 3D simulations on the upward overshooting in turbulent compressible convection. With this simulated 3D data, we calibrate the convective, diffusive, and isotropic coefficients for Xiong's 1D Reynolds stress model. We calibrate the convective and isotropic coefficients ($c_{1,.}$ and $c_{3}$) with the data in the convectively unstable zone, and the diffusive coefficients ($c_{2,.}$) with the data in the convectively stable zone, respectively. It has been found that the calibrated coefficients $c_{1,.}$ and $c_{3}$ are close to those calibrated by the 3D data of the simulations of the pure convection zone \citep{cai2018numerical}. However, the calibrated diffusive coefficients $c_{2,.}$ deviate significantly from those calibrated in \citet{cai2018numerical}. As \citet{cai2018numerical} calibrated $c_{2,.}$ by the boundary effect, we suspect that the diffusive effect induced by the upper boundary is stronger than by the adjacent stable zone. With the 3D data, we have checked the validity of the downgradient approximations. We find that the prediction of the downgradient approximations on the third-order moments is unsatisfactory in the convectively unstable zone. The prediction on the features in convectively stable zone, such as the dips and bumps, is much better. Although the TOMs differ significantly from the DGAs in the convectively unstable zone, the difference between their derivatives is diminished. In Xiong's 1D RSM, fortunately only the derivatives of TOMs are involved in the turbulent moments equations. For this reason, the performance of 1D RSM is reasonable in the application to the real stars \citep{xiong2001structure,xiong2010non}.

Including the equations on thermal structures, we have solved the full set of Xiong's 1D nonlocal turbulent equations with the calibrated coefficients. We find that the DGAs have better performance in the prediction of the second-order moments. Some features like the bumps and dips in the overshooting zone are well captured by the model. Most importantly, the Reynolds stress model has successfully produced the thermal adjustment layer and turbulent dissipation layer, which were identified in the previous 3D simulations \citep{cai2020upward}. Comparing the 1D and 3D results, we have found two useful indicators on measuring the extent of overshooting distance: the inflection point of $\langle\overline{\theta^2}\rangle$ (close to the boundary of the thermal adjustment layer), and the location point of ${\rm Pe}_{Hp}=1$ (close to the boundary of the turbulent dissipation layer). Apart from these similarities, there are also some differences between the 1D and 3D results. The 1D RSM predicts a lower convective flux (and negligible turbulent kinetic energy flux) in the convection zone. In addition, the temperature gradient of 1D case turns to be subadiabatic below the interface, contrary to the superadibatic temperature gradient obtained in 3D simulations. Subadiabatic temperature gradient in the convection zone has been observed in the 3D simulations of the overshooting in turbulent compressible convection \citep{chan1992downflows,kapyla2017extended,hotta2017solar}. It seems that the subadiabatic temperature gradient prefers to appear at the bottom of the convection zone. In the absence of the convectively stable zone, \citet{cai2018numerical} also observed the subadiabatic temperature gradient near the bottom of the convection zone. Recently, \citet{korre2017weakly} has observed that the subadiabatic temperature gradient occurs at the top of the convection zone in their simulations of the weakly compressible convection (without adjacent stable zone) in a spherical shell. This result is different from the other findings. Since both the degree of compressibility and geometrical shape can affect the result, it remains unclear which effect causes this difference. Identifying the reason requires more explorations on the parameter space. So far, both our numerical simulations and the 1D nonlocal model have not considered the effect of rotation. Rotation has important effect on convection and overshooting. For example, penetration depth may vary with latitudes because the Coriolis effect differs at high and low latitudes \citep{browning2004simulations,pal2008turbulent}. In certain circumstance, vortices might appear when the Rossby number is small \citep{kapyla2011starspots}. In even more extreme rotation rates, the spherical shape of the star can be deformed by the strong centrifugal force. Investigation of the rotational effect is beyond the scope of this paper. We plan to conduct this research in the future.

\acknowledgements
I thank D.R. Xiong for the helpful discussion on his turbulent convection model. I was financially supported by NSFC (Nos. 11503097,11521101), the Guangdong Basic and Applied Basic Research Foundation (No. 2019A1515011625), the Science and Technology Program of Guangzhou (No. 201707010006), the Science and Technology Development Fund, Macau SAR (No. 0045/2018/AFJ), and the Independent Innovation Project of China Academy of Space Technology. The simulations were performed on the supercomputers at the Purple Mountain Observatory, and the National Supercomputer Center in Guangzhou.




\bibliographystyle{aasjournal}
\bibliography{ov3}



\end{document}